\documentstyle[12pt]{article}
\newlength{\extraspace}
\newlength{\extraspaces}
\catcode`\@=11
\def\numberbysection{\@addtoreset{equation}{section}
\def\theequation{\arabic{section}.\arabic{equation}}}
\newcommand{\be}{\begin{equation}
\addtolength{\abovedisplayskip}{\extraspaces}
\addtolength{\belowdisplayskip}{\extraspaces}
\addtolength{\abovedisplayshortskip}{\extraspace}
\addtolength{\belowdisplayshortskip}{\extraspace}}
\newcommand{\ee}{\end{equation}}
\newcommand{\ba}{\begin{eqnarray}
\addtolength{\abovedisplayskip}{\extraspaces}
\addtolength{\belowdisplayskip}{\extraspaces}
\addtolength{\abovedisplayshortskip}{\extraspace}
\addtolength{\belowdisplayshortskip}{\extraspace}}
\newcommand{\ea}{\end{eqnarray}}
\newcommand{\newsection}[1]{
\vspace{7mm}
\pagebreak[3]
\addtocounter{section}{1}
\setcounter{equation}{0}
\setcounter{subsection}{0}
\setcounter{footnote}{0}
\begin{center}
{\large {\bf \thesection. #1}}
\end{center}
\nopagebreak
\medskip
\nopagebreak
\hspace{3mm}}
\newcommand{\nonu}{\nonumber \\[.5mm]}
\newcommand{\A}{&\!\!\!}

\newcommand{\C}{{\bf C}}
\newcommand{\R}{{\bf R}}
\newcommand{\tr}{\, {\rm tr}}
\newcommand{\e}{\, {\rm e}}

\newcommand{\ket}[1]{\left\vert {#1} \right\rangle}

\begin{document}
\addtolength{\baselineskip}{.7mm}
\thispagestyle{empty}
\begin{flushright}
STUPP--98--146 \\
{\tt hep-th/9802138} \\ 
February, 1998
\end{flushright}
\vspace{15mm}
\begin{center}
{\Large{\bf Introduction to Supergravities \\[2mm] 
in Diverse Dimensions\footnote{An expanded version of 
a review talk at YITP workshop on Supersymmetry, 27 -- 30 March, 1996}
}} \\[30mm]
{\sc Yoshiaki Tanii}\footnote{
\tt e-mail: tanii@th.phy.saitama-u.ac.jp} \\[7mm] 
{\it Physics Department, Faculty of Science \\
Saitama University, Urawa, Saitama 338, Japan} \\[30mm]
{\bf Abstract}\\[5mm]
{\parbox{13cm}{\hspace{5mm}
Supergravities in four and higher dimensions are reviewed. 
We discuss the action and its local symmetries of $N=1$ 
supergravity in four dimensions, possible types of spinors 
in various dimensions, field contents of supergravity multiplets, 
non-compact bosonic symmetries, non-linear sigma models, 
duality symmetries of antisymmetric tensor fields and super 
$p$-branes. 
}}
\end{center}
\vfill
\newpage
\setcounter{section}{0}
\setcounter{equation}{0}
\numberbysection
%
%
\newsection{Introduction}
Recently interest in supergravities in various space-time 
dimensions has been much increased due to their relevance 
to string dualities \cite{HTW}. 
String dualities sometimes relate string theories at strong 
coupling and those at weak coupling, and are extremely useful 
to understand non-perturbative properties of string theories. 
However, at present understanding of string theories, 
it is difficult to show dualities directly in full 
string theories. Massless sectors of superstring theories 
are described by supergravities, for which complete field 
theoretic formulations are known at the classical level. 
One may try to obtain information about string 
dualities by studying supergravities. 
\par
One of the purposes of this paper is to explain the basic 
ideas of supergravities to those who are not familiar with 
supergravities. We try to make the discussions pedagogical 
and often present explicit calculations. 
Another purpose is to collect relevant formulae together 
in one place, which may be useful when one discusses string 
dualities. They include possible types of spinors in various 
dimensions, field contents of supergravity multiplets, 
bosonic symmetries of supergravities, etc. 
We do not discuss quantum properties of supergravities 
such as ultraviolet divergences or anomalies. 
\par
Supergravities are field theories which have the local 
supersymmetry. A transformation parameter of the rigid 
supersymmetry is a constant spinor $\epsilon^\alpha$. 
(For a review of the rigid supersymmetry see 
ref.\ \cite{WB}, for instance.) 
To construct theories which have the local supersymmetry 
we introduce a gauge field $\psi_\mu^\alpha(x)$, 
which has a vector index $\mu$ in addition to the spinor 
index $\alpha$. The transformation law of the local 
supersymmetry is 
$\delta_Q \psi_\mu^\alpha(x) 
= \partial_\mu \epsilon^\alpha + \cdots$, 
where the transformation parameter $\epsilon^\alpha(x)$ 
is an arbitrary function of the space-time coordinates 
$x^\mu$. Such a field $\psi_\mu^\alpha(x)$ is the 
Rarita-Schwinger field representing a spin ${3 \over 2}$ 
particle. However, that is not all we need. The 
anticommutation relation of supercharges $Q^\alpha$ produces 
the translation generators $P_a$: 
\be
\{ Q^\alpha, \bar Q_\beta \} 
= (\gamma^a)^\alpha{}_\beta P_a. 
\ee
Therefore, 
we expect that gauging of supersymmetry leads to gauging of 
translation. Since the local translation is the general 
coordinate transformation, we also need the gravitational 
field $g_{\mu\nu}(x)$ as a gauge field. 
To summarize, supergravities are theories, which 
are invariant under the local supersymmetry transformation 
as well as the general coordinate transformation. 
They contain the gravitational field $g_{\mu\nu}(x)$ and 
the Rarita-Schwinger field $\psi_\mu^\alpha(x)$. 
\par
In the next section we discuss supergravities in four 
dimensions in some detail. 
To generalize these results to higher dimensions 
we first discuss what types of spinor representations 
are possible in general dimensions in sect.\ 3. 
Then, possible superalgebras and supergravity multiplets 
in higher dimensions are given in sect.\ 4. 
Properties of higher dimensional supergravities are 
discussed in sect.\ 5. 
In sects.\ 6 and 7 we discuss subjects related to bosonic 
non-compact symmetries appearing in supergravities. 
In sect.\ 6 we explain how scalar fields are described by 
G/H non-linear sigma models with a non-compact Lie group G 
and its maximal compact subgroup H. 
In sect.\ 7 we discuss duality symmetries, which transform 
field strength of antisymmetric tensor fields into their 
duals. The non-compact group G acts on the antisymmetric 
tensor fields as duality transformations. 
Finally, in sect.\ 8 we briefly discuss super $p$-branes 
($p$-dimensionally extended objects), which are closely 
related to supergravities.  
The vielbein formulation of gravity is summarized 
in Appendix A. 
A proof of the local supersymmetry invariance of $N=1$ 
supergravity in four dimensions is given in Appendix B. 
We have not tried to give complete references to 
the original papers. 
For more complete references see ref.\ \cite{SS}. 
Other useful review papers on supergravities are 
refs.\ \cite{vN}, \cite{SCHERK}. 
%
%
\newsection{Supergravities in four dimensions}
In this section we shall consider supergravities in $d=4$ 
space-time dimensions. Massless irreducible 
representations of the $N=1$ superalgebra consist of two 
states with helicities differing by ${1 \over 2}$. 
In particular, we have representations of helicities 
$(2, {3 \over 2})$ and $(-{3 \over 2}, -2)$, which correspond 
to a pair of fields $(g_{\mu\nu}(x), \psi_\mu^\alpha(x))$. 
Therefore, there is a possibility of constructing a supergravity 
which contains only these two fields. Such a theory was 
explicitly constructed in ref.\ \cite{FvNF}. 
\par
The field content of the $d=4$, $N=1$ supergravity is 
the vierbein (tetrad) $e_\mu{}^a(x)$ and a Majorana 
Rarita-Schwinger field $\psi_\mu(x)$. 
The vierbein is related to the metric as 
$g_{\mu\nu} = e_\mu{}^a e_\nu{}^b \eta_{ab}$, where 
$\eta_{ab}= {\rm diag}(+1, -1, -1, -1)$ is the flat Minkowski 
metric. The vielbein formalism of gravity is reviewed 
in Appendix A. 
The Rarita-Schwinger field satisfies the Majorana 
condition $\psi_\mu^c (\equiv C \bar\psi_\mu^T) = \psi_\mu$, 
where $C$ is the charge conjugation matrix satisfying 
\be
C^{-1} \gamma^a C = - \gamma^{aT}, \qquad
C^T = - C. 
\ee
The Lagrangian consists of the Einstein term and 
the Rarita-Schwinger term 
\be
{\cal L} = - {1 \over 4} e \hat R - {1 \over 2} i e 
\bar\psi_\mu \gamma^{\mu\nu\rho} \hat D_\nu \psi_\rho, 
\label{4daction}
\ee
where $e = \det e_\mu{}^a$ and $\gamma$'s with multiple indices 
are antisymmetrized products of gamma matrices with unit strength 
\be
\gamma^{\mu\nu\rho} = {1 \over 3!} \left( 
\gamma^\mu \gamma^\nu \gamma^\rho 
\pm \mbox{permutations of $\mu\nu\rho$} \right). 
\label{gammaproduct}
\ee
The curvature and the covariant derivative are defined by 
\ba
\hat R \A = \A e_a{}^\mu e_b{}^\nu \hat R_{\mu\nu}{}^{ab}, \nonu
\hat R_{\mu\nu}{}^{ab} \A = \A \partial_\mu \hat\omega_\nu{}^{ab} 
- \partial_\nu \hat\omega_\mu{}^{ab} 
+ \hat\omega_\mu{}^a{}_c \hat\omega_\nu{}^{cb} 
- \hat\omega_\nu{}^a{}_c \hat\omega_\mu{}^{cb}, \nonu
\hat D_\nu \psi_\rho \A = \A \left( \partial_\nu 
+ {1 \over 4} \hat\omega_\nu{}^{ab} 
\gamma_{ab} \right) \psi_\rho. 
\ea
The spin connection $\hat\omega_\mu{}^{ab}$ used here is given by 
\be
\hat\omega_{\mu ab} = \omega_{\mu ab} 
- {1 \over 2} i \bar\psi_a \gamma_\mu \psi_b 
- {1 \over 2} i \bar\psi_\mu \gamma_a \psi_b 
+ {1 \over 2} i \bar\psi_\mu \gamma_b \psi_a, 
\label{sct}
\ee
where $\omega_{\mu ab}$ is the spin connection without 
torsion given in eq.\ (\ref{a14}). 
The spin connection (\ref{sct}) has a torsion 
depending on the Rarita-Schwinger field 
\be
\hat D_\mu e_\nu{}^a - \hat D_\nu e_\mu{}^a 
= - i \bar\psi_\mu \gamma^a \psi_\nu. 
\label{torsion}
\ee
If one wishes, it is also possible to express the 
Lagrangian using the torsionless spin connection 
$\omega_{\mu ab}$ but with explicit 4-fermi terms 
\be
{\cal L} = - {1 \over 4} e R - {1 \over 2} i e 
\bar\psi_\mu \gamma^{\mu\nu\rho} D_\nu \psi_\rho 
+ (\mbox{4-fermi terms}), 
\label{4daction2}
\ee
where $R$ and $D_\nu$ are defined by using the 
torsionless spin connection. 
\par
The Lagrangian (\ref{4daction}) is invariant under three 
kinds of local symmetries up to total divergences: 

\noindent (i) general coordinate transformations 
\ba
\delta_G(\xi) e_\mu{}^a \A = \A - \xi^\nu \partial_\nu e_\mu{}^a 
- \partial_\mu \xi^\nu e_\nu{}^a, \nonu
\delta_G(\xi) \psi_\mu \A = \A - \xi^\nu \partial_\nu \psi_\mu 
- \partial_\mu \xi^\nu \psi_\nu, 
\ea

\noindent (ii) local Lorentz transformations 
\ba
\delta_L(\lambda) e_\mu{}^a \A = \A - \lambda^a{}_b e_\mu{}^b, \nonu
\delta_L(\lambda) \psi_\mu \A = \A 
- {1 \over 4} \lambda^{ab} \gamma_{ab} \psi_\mu, 
\ea

\noindent (iii) local supertransformations 
\ba
\delta_Q(\epsilon) e_\mu{}^a \A = \A 
- i \bar\epsilon \gamma^a \psi_\mu, \nonu
\delta_Q(\epsilon) \psi_\mu \A = \A \hat D_\mu \epsilon 
\equiv \left( \partial_\mu + {1 \over 4} \hat\omega_\mu{}^{ab} 
\gamma_{ab} \right) \epsilon, 
\label{localsusy}
\ea

\noindent
where the transformation parameters $\xi^\mu(x)$, 
$\lambda^a{}_b(x)$ ($\lambda^{ab} = - \lambda^{ba}$) 
and $\epsilon_\alpha(x)$ ($\epsilon^c = \epsilon$) are 
arbitrary functions of the space-time coordinates $x^\mu$. 
The invariance under the bosonic transformations (i), (ii) is 
manifest. The invariance under the local 
supertransformations (iii) is shown in Appendix B. 
\par
These local transformations satisfy the following 
commutator algebra: 
\ba
[ \delta_G(\xi_1), \delta_G(\xi_2) ] 
\A = \A \delta_G(\xi_1 \cdot \partial \xi_2 
- \xi_2 \cdot \partial \xi_1), \nonu
[ \delta_L(\lambda_1), \delta_L(\lambda_2) ] 
\A = \A \delta_L([\lambda_1, \lambda_2]), \nonu
[ \delta_G(\xi), \delta_L(\lambda) ] 
\A = \A \delta_L(\xi \cdot \partial \lambda), \nonu
[ \delta_G(\xi), \delta_Q(\epsilon) ] 
\A = \A \delta_Q(\xi \cdot \partial\epsilon), \nonu
[ \delta_L(\lambda), \delta_Q(\epsilon) ] 
\A = \A \delta_Q({\textstyle {1 \over 4}} \lambda^{ab} 
\gamma_{ab} \epsilon), \nonu
[ \delta_Q(\epsilon_1), \delta_Q(\epsilon_2) ] 
\A = \A \delta_G(\xi) + \delta_L(\xi \cdot \hat\omega) 
+ \delta_Q(\xi \cdot \psi), \qquad 
\xi^\mu = i \bar\epsilon_2 \gamma^\mu \epsilon_1. 
\label{commutatoralg}
\ea
These commutation relations except the last one can be easily 
shown. The last commutation relation is shown in Appendix B. 
To obtain the last commutation relation one has to use 
field equations derived from the Lagrangian (\ref{4daction}). 
In this sense the algebra closes only on-shell. 
In the present theory it is possible to close the commutator 
algebra off-shell by introducing an appropriate set of 
auxiliary fields, which have no dynamical degrees of freedom. 
Theories with off-shell algebra are more convenient, 
although not indispensable, when one fixes a gauge of local 
symmetries and when one couples matter supermultiplets. 
For general supergravities 
(those with extended supersymmetry and/or in higher 
dimensions) such an off-shell formulation is not known. 
\par
One can couple matter supermultiplets to the supergravity 
multiplet ($e_\mu{}^a$, $\psi_\mu$). There are two kinds of 
matter supermultiplets in the $d=4$, $N=1$ supersymmetry. 
A chiral multiplet ($\phi$, $\lambda$) consists of 
a complex scalar field $\phi(x)$ and a Majorana spinor 
field $\lambda(x)$. A vector multiplet 
($A_\mu$, $\chi$) consists of a vector field 
$A_\mu(x)$ and a Majorana spinor field $\chi(x)$. 
The Lagrangian and the supertransformations 
of matter coupled theories can be obtained by using either of 
the Noether method \cite{vN}, the superspace 
formulation \cite{WB}, \cite{GGRS} or the tensor 
calculus \cite{KU}. In this paper, however, we do not discuss 
such matter couplings but concentrate on pure supergravities, 
which contain only supergravity multiplets. 
\par
So far we have considered the $N=1$ supergravity based on 
the $N=1$ supersymmetry. We can also consider a gauging of 
the $N$-extended supersymmetry with $N$ transformation 
parameters $\epsilon^i$ ($i=1, 2, \cdots, N$) \cite{SCHERK}. 
We need $N$ gravitinos $\psi^i_\mu(x)$ 
($i = 1, 2, \cdots, N$), which transform as 
$\delta_Q \psi_\mu^i = \partial_\mu \epsilon^i + \cdots$. 
To make supermultiplets of the extended supersymmetry 
we also need other fields in addition to the metric and 
the Rarita-Schwinger fields. Supergravity multiplets of 
extended supersymmetries are listed 
in Table \ref{tablethree} of sect.\ 4. 
Representations of $N \ge 9$ supersymmetry algebra contain 
particles with helicities greater than two. Since consistent 
interacting theories of particles with such high helicities 
are not known, $N \ge 9$ supergravities have not 
been constructed. 
\par
$N$-extended supergravities contain ${1 \over 2}N(N-1)$ vector 
fields denoted $B_\mu$ in Table \ref{tablethree}. 
They are ${\rm U}(1)^N$ gauge fields. 
It is possible to construct theories in which the vector fields 
are O($N$) non-abelian gauge fields. Such theories are called 
gauged supergravities \cite{FD}. 
Their Lagrangians contain a cosmological 
term $e$ proportional to $g^2$ and a gravitino mass term 
$\bar\psi_\mu \gamma^{\mu\nu} \psi_\nu$ 
proportional to $g$, where $g$ is a gauge coupling constant 
of the non-abelian gauge fields. 
In the limit $g \rightarrow 0$ they reduce 
to the ordinary supergravities. 
The $N = 1$ theory also has a generalization with a cosmological 
term and a gravitino mass term. 
%
%
\newsection{Spinors in higher dimensions}
To construct supergravities in higher dimensions we need to 
know what kinds of spinors we can define in each 
dimension \cite{GSO}. We consider spinor representations of 
the group SO($t$,$s$) with an invariant metric 
\be
\eta_{ab} = {\rm diag} (\underbrace{+1, \cdots, +1}_t, 
\underbrace{-1, \cdots, -1}_s), \qquad d = t+s. 
\ee
The $d$-dimensional Minkowski space-time corresponds to 
the case $t=1$, $s=d-1$. 
Gamma matrices $\gamma^a$ ($a = 1, 2, \cdots, d$) of 
SO($t$, $s$) satisfy the anticommutation relation 
\be
\{ \gamma^a, \gamma^b \} = 2 \eta^{ab}. 
\ee
Matrices $\gamma^a$ are hermitian for $a = 1, \cdots, t$ 
and antihermitian for $a = t+1, \cdots, d$. 
The smallest matrices satisfying this anticommutation relation are 
$2^{\left[{d \over 2}\right]} \times 2^{\left[{d \over 2}\right]}$, 
where $[x]$ is the largest integer not larger than $x$. 
An explicit representation of the gamma matrices can be 
constructed as tensor products of $2 \times 2$ matrices. 
In even dimensions (e.g., $t = 2n$, $s = 0$) we can use 
the following tensor products of $n$ $2 \times 2$ matrices: 
\ba
\gamma^1 \A = \A \sigma^1 \otimes {\bf 1} \otimes 
\cdots \otimes {\bf 1}, \nonu
\gamma^2 \A = \A \sigma^2 \otimes {\bf 1} \otimes 
\cdots \otimes {\bf 1}, \nonu
\gamma^3 \A = \A \sigma^3 \otimes \sigma^1 \otimes {\bf 1} 
\otimes \cdots \otimes {\bf 1}, \nonu
\gamma^4 \A = \A \sigma^3 \otimes \sigma^2 \otimes {\bf 1} 
\otimes \cdots \otimes {\bf 1}, \nonu
\A \vdots \A \nonu
\gamma^{2k+1} \A = \A 
\underbrace{\sigma^3 \otimes \cdots \otimes \sigma^3}_k \otimes 
\sigma^1 \otimes {\bf 1} \otimes \cdots \otimes {\bf 1}, \nonu
\gamma^{2k+2} \A = \A 
\underbrace{\sigma^3 \otimes \cdots \otimes \sigma^3}_k \otimes 
\sigma^2 \otimes {\bf 1} \otimes \cdots \otimes {\bf 1}, \nonu
\A \vdots \A \nonu
\gamma^{2n-1} \A = \A \sigma^3 \otimes \cdots \otimes \sigma^3 
\otimes \sigma^1, \nonu
\gamma^{2n} \A = \A \sigma^3 \otimes \cdots \otimes \sigma^3 
\otimes \sigma^2, 
\ea
where {\bf 1} is the $2 \times 2$ unit matrix and 
$\sigma^i$ ($i=1,2,3$) are the Pauli matrices. 
Gamma matrices in odd dimensions ($d=t+s=2n+1$) can be constructed 
by using those in even ($d=t+(s-1)=2n$) dimensions. 
We can use gamma matrices $\gamma^a$ ($a=1, \cdots, 2n$) 
of SO($t$, $s-1$) for the first $2n$ gamma matrices 
of SO($t$, $s$). The last matrix $\gamma^{2n+1}$ can be taken as 
\be
\gamma^{2n+1} = (-1)^{{1 \over 4}(s-t)} \, i \, \gamma^1 \gamma^2 
\cdots \gamma^{2n}. 
\ee
\par
Spinors of SO($t$,$s$) have $2^{\left[{d \over 2}\right]}$ 
complex components in general and transform under the 
Lorentz transformation as 
\be
\delta_L \psi = - {1 \over 4} \lambda^{ab} \gamma_{ab} \psi. 
\label{spinorlorentz}
\ee
As we will discuss below, we can reduce the number of 
independent components of spinors by imposing 
Weyl and/or Majorana conditions. 
These conditions must be consistent with the Lorentz 
transformation law (\ref{spinorlorentz}). 
Spinors satisfying these conditions are called Weyl spinors 
and Majorana spinors respectively. 
General spinors without any condition are called Dirac spinors. 
To discuss supersymmetry it is convenient to use spinors with 
the smallest number of independent components in each dimension. 
\par
Weyl spinors are those having a definite chirality in even 
dimensions. We define the chirality of spinors as an eigenvalue 
of the matrix 
\be
\bar\gamma 
= (-1)^{{1 \over 4}(s-t)} \gamma^1 \gamma^2 \cdots \gamma^d 
\label{chirality}
\ee
satisfying 
\be
{\bar\gamma}^2 = 1, \qquad \{ \bar\gamma, \gamma^a \} = 0. 
\ee
This matrix $\bar\gamma$ is a generalization of $\gamma_5$ 
in four dimensions. Weyl spinors with positive (or negative) 
chirality are defined by 
\be
\bar\gamma \psi = \psi \qquad 
\left( {\rm or}\ \ \bar\gamma \psi = -\psi \right). 
\ee
It is easy to see that these conditions are consistent with 
eq.\ (\ref{spinorlorentz}). 
The matrix $\bar\gamma$ and therefore Weyl spinors 
can be defined in any even dimensions. 
In odd dimensions $\bar\gamma$ is proportional 
to the unit matrix and one cannot define Weyl spinors. 
\par
Majorana spinors are those satisfying a certain kind of 
reality condition. The Majorana condition is 
\be
\psi^c = \psi, 
\label{majorana}
\ee
where the superscript ${}^c$ represents a charge conjugation. 
We shall discuss the charge conjugation for $d=t+s=$ even and 
for $d=t+s=$ odd separately. 
\par
Let us first consider the case $d=t+s=$ even. 
The matrices $\pm (\gamma^a)^*$ satisfy the same anticommutation 
relation as $\gamma^a$. Then, it can be shown that there 
exist matrices $B_+$ and $B_-$, which relate $\pm (\gamma^a)^*$ 
to $\gamma^a$ by similarity transformations 
\ba
(\gamma^a)^* \A = \A B_+ \gamma^a B_+^{-1}, \nonu
-(\gamma^a)^* \A = \A B_- \gamma^a B_-^{-1}. 
\ea
The charge conjugation is defined by using one of 
these matrices as 
\be
\psi^c = B_+^{-1} \psi^* \qquad {\rm or} \qquad 
\psi^c = B_-^{-1} \psi^*. 
\label{chargeconj}
\ee
It can be shown that $B_\pm$ satisfy 
\be
B_\pm^* B_\pm = \epsilon_\pm(t,s) \; {\bf 1}, \qquad 
\epsilon_\pm(t,s) 
= \sqrt{2} \cos \left[ {\pi \over 4}(s-t \pm 1) \right]. 
\ee
We have summarized the values of $\epsilon_\pm$ in 
Table \ref{tableone}. 
\begin{table}[tbp]
\begin{center}
\begin{tabular}{|c|c|c|c|c|c|c|c|c|} \hline
\multicolumn{1}{|c|}{$s-t$} &
\multicolumn{1}{|c|}{1} &
\multicolumn{1}{|c|}{2} &
\multicolumn{1}{|c|}{3} &
\multicolumn{1}{|c|}{4} &
\multicolumn{1}{|c|}{5} &
\multicolumn{1}{|c|}{6} &
\multicolumn{1}{|c|}{7} &
\multicolumn{1}{|c|}{8} \\ \hline
$\epsilon_+$ & No & $-1$ & $-1$ & $-1$ & No & 
$+1$ & $+1$ & $+1$ \\ \hline
$\epsilon_-$ & $+1$ & $+1$ & No & $-1$ & $-1$ & 
$-1$ & No & $+1$ \\ \hline
\end{tabular}
\end{center}
\caption{The values of $\epsilon_\pm$.}
\label{tableone}
\end{table}
\par
The reason why this operation is called the charge conjugation 
can be seen as follows. 
When a spinor field $\psi$ satisfies the Dirac equation in the 
presence of an electromagnetic gauge field $A_\mu$ 
\be
( i \gamma^\mu \partial_\mu - e \gamma^\mu A_\mu - m ) \psi = 0, 
\ee
the charge conjugated field $\psi^c$ in 
eq.\ (\ref{chargeconj}) satisfies 
\be
( i \gamma^\mu \partial_\mu + e \gamma^\mu A_\mu \pm m ) \psi^c = 0. 
\ee
Thus, the sign in front of the charge $e$ has changed. 
In the mass term the upper sign $+$ is for $B_+$ and the 
lower $-$ is for $B_-$. 
The definition of the charge conjugation in 
eq.\ (\ref{chargeconj}) is equivalent to the usual one using 
the charge conjugation matrix $C$. To show this we introduce 
the Dirac conjugate of $\psi$ as 
\be
\bar\psi = \psi^\dagger A, \qquad 
A = \gamma^1 \gamma^2 \cdots \gamma^t. 
\ee
Then, eq.\ (\ref{chargeconj}) can be rewritten as 
\be
\psi^c = C_+ \bar\psi^T \qquad {\rm or} \qquad
\psi^c = C_- \bar\psi^T \qquad 
\left( C_\pm = B_\pm^{-1} A^{-1 T} \right). 
\ee
The charge conjugation matrices $C_\pm$ satisfy 
\ba
\gamma^{aT} \A = \A \pm (-1)^{t+1} C_\pm^{-1} \gamma^a C_\pm, \nonu
C_\pm^\dagger C_\pm \A = \A {\bf 1}, \qquad
C_\pm^T = (\pm 1)^t (-1)^{{1 \over 2}t(t-1)} \epsilon_\pm C_\pm. 
\ea
The usual four-dimensional charge conjugation matrix $C$ 
used in sect.\ 2 is $C_-$. 
\par
For the Majorana condition (\ref{majorana}) to be consistent 
we must have $(\psi^c)^c = \psi$, which is equivalent to 
$B_+^* B_+ = 1$ or $B_-^* B_- = 1$. Therefore, for $t+s=$ even, 
Majorana spinors can be defined only when $\epsilon_+(t,s) = 1$ 
or $\epsilon_-(t,s) = 1$. 
Sometimes, spinors satisfying eq.\ (\ref{majorana}) with the 
charge conjugation defined by using the matrix $B_+$ are called 
pseudo Majorana spinors, while those using $B_-$ are called 
Majorana spinors. From Table \ref{tableone} we can see in which 
dimensions (pseudo) Majorana spinors can be defined. 
\par
The charge conjugation for $d=t+s=$ odd is defined by 
using the matrices $B_\pm$ used in even dimensions. 
Recall that gamma matrices in $d$ (odd) dimensions can be 
constructed from those in $d-1$ (even) dimensions. 
The first $d-1$ matrices $\gamma^a$ ($a=1, 2, \cdots, d-1$) 
are taken to be those of $d-1$ dimensions. 
The last matrix $\gamma^d$ is taken to be $i \bar\gamma$ 
($\bar\gamma$) if $a=d$ is a space-like (time-like) direction. 
Then the matrices $B_\pm$ used in $d-1$ dimensions satisfy 
\ba
B_\pm \gamma^a B_\pm^{-1} \A = \A \pm (\gamma^a)^* 
\qquad (a = 1, \cdots, d-1), \nonu
B_\pm \gamma^d B_\pm^{-1} \A = \A (-1)^{{1 \over 2}(s-t+1)} 
(\gamma^d)^*. 
\ea
When $(-1)^{{1 \over 2}(s-t+1)} = 1$ 
($(-1)^{{1 \over 2}(s-t+1)} = -1$), 
the signs on the right hand side are the same for all $\gamma^a$ 
($a = 1, \cdots, d$), and we can use $B_+$ ($B_-$) to define 
the charge conjugation. 
As in the case $d=$ even, the charge conjugation must satisfy 
$(\psi^c)^c = \psi$ to define (pseudo) Majorana spinors. 
Possible $B_\pm$ and corresponding $\epsilon_\pm$ are listed 
in Table \ref{tableone}. 
\par
We can also define (pseudo) Majorana-Weyl spinors, 
which satisfy both of the (pseudo) Majorana condition 
$\psi^c = \psi$ and the Weyl condition 
$\bar\gamma \psi = \psi$ (or $\bar\gamma \psi = -\psi$). 
(Pseudo) Majorana-Weyl spinors are possible only if these two 
conditions are consistent, i.e., $\psi$ and $\psi^c$ have the 
same chirality. 
In general, when a spinor $\psi$ has a chirality $+$ ($-$), 
the charge conjugated spinor $\psi^c$ has a chirality 
$(-1)^{{1 \over 2}(s-t)}$ ($-(-1)^{{1 \over 2}(s-t)}$). 
Therefore, (pseudo) Majorana-Weyl spinors can be defined only 
when $s-t = 0$ mod 4. In particular, they can be defined in 
$d = 2$ mod 4 for Minkowski signature $t=1$, $s=d-1$. 
\par
Possible types of spinors in various dimensions with Minkowski 
signature $t=1$, $s=d-1$ are summarized in Table \ref{tabletwo}. 
This table is periodic in dimensions $d$ with a period 8. 
\begin{table}[tbp]
\begin{center}
\begin{tabular}{|r|c|c|c|c|c|} \hline
\multicolumn{1}{|c|}{$d$} &
\multicolumn{1}{|c|}{W} &
\multicolumn{1}{|c|}{M} &
\multicolumn{1}{|c|}{pM} &
\multicolumn{1}{|c|}{MW} &
\multicolumn{1}{|c|}{pMW} \\ \hline
2  & $\circ$ & $\circ$ & $\circ$ & $\circ$ & $\circ$ \\ 
3  &         & $\circ$ &         &         &         \\ 
4  & $\circ$ & $\circ$ &         &         &         \\ 
5  &         &         &         &         &         \\ 
6  & $\circ$ &         &         &         &         \\ 
7  & \phantom{pMW} & \phantom{pMW} & \phantom{pMW} & \phantom{pMW} 
   & \phantom{pMW} \\ 
8  & $\circ$ &         & $\circ$ &         &         \\ 
9  &         &         & $\circ$ &         &         \\ 
10 & $\circ$ & $\circ$ & $\circ$ & $\circ$ & $\circ$ \\ 
11 &         & $\circ$ &         &         &         \\ \hline
\end{tabular}
\end{center}
\caption{Possible types of spinors in $d$-dimensional Minkowski 
space-time ($t=1$, $s=d-1$). W, M, pM, MW, pMW denote Weyl, 
Majorana, pseudo Majorana, Majorana-Weyl and pseudo 
Majorana-Weyl spinors respectively.}
\label{tabletwo}
\end{table}
\par
When $(\psi^c)^c = -\psi$, we cannot impose the (pseudo) 
Majorana condition $\psi^c = \psi$ and we have to use Dirac 
spinors (or Weyl spinors in even dimensions). Alternatively, we can 
introduce even numbers of spinors $\psi^i$ ($i = 1, 2, \cdots, 2n$) 
and impose the condition 
\be
\psi^i = \Omega^{ij} (\psi^j)^c, 
\ee
where $\Omega^{ij} = - \Omega^{ji}$ is a constant antisymmetric 
matrix. Spinors satisfying such a condition are called symplectic 
(pseudo) Majorana spinors. $2n$ symplectic (pseudo) Majorana spinors 
are equivalent to $n$ Dirac spinors. 
Sometimes it is more convenient to use symplectic (pseudo) Majorana 
spinors than Dirac spinors, especially when the theory has a 
symplectic symmetry. 
%
%
\newsection{Superalgebras and supergravity multiplets}
Field contents of supergravities are determined by irreducible 
representations of the super Poincar\'e algebras \cite{STRATHDEE}. 
The super Poincar\'e algebras consist of 
generators of translations $P_a$, 
generators of Lorentz transformations $M_{ab}$, 
supercharges $Q^{\alpha i}$, generators of automorphism group $T^A$ 
and ``central'' charges $Z^{ij}$. 
Nonvanishing (anti) commutation relations besides $\{ Q, Q \}$ 
are, in addition to the usual commutators of the 
Poincar\'e algebra, 
\ba
[ M_{ab}, Q^i ] \A = \A {1 \over 2} \gamma_{ab} Q^i, \qquad
[ T^A, Q^i ] = (t^A)^i{}_j Q^j, \nonu
[ T^A, Z^{ij} ] \A = \A (t^A)^i{}_k Z^{kj} 
+ (t^A)^j{}_k Z^{ik}, \qquad
[ T^A, T^B ] = f^{AB}{}_C T^C, 
\ea
where $t^A$ and $f^{AB}{}_C$ are representation matrices and 
the structure constant of the Lie algebra of the automorphism group. 
\par
The automorphism group K and the form of anticommutators 
$\{ Q, Q \}$ depend on the spinor type of $Q^i$. 

\noindent (a) $d = 4, 8$ mod 8 

\noindent 
The supercharges are Weyl spinors with positive chirality 
$Q_+^i$ ($i = 1, 2, \cdots, N$). Their charge conjugations 
have negative chirality $(Q_+^i)^c = Q_{- \, i}$, where 
the charge conjugation matrix $C = C_-$ ($C = C_+$) is used 
for $d=4$ ($d=8$) mod 8. The automorphism group is K = U($N$). 
Anticommutators of the supercharges are 
\ba
\{Q_+^i, Q_{-j}^T \} \A = \A {1 \over 2} 
(1+\bar\gamma) \gamma^a C P_a \delta_j^i, \nonu
\{Q_+^i, Q_+^{jT} \} \A = \A {1 \over 2} 
(1+\bar\gamma) C Z^{ij}, 
\ea
where $Z^{ij} = - Z^{ji}$ for $d=4$ mod 8 and $Z^{ij} = Z^{ji}$ 
for $d=8$ mod 8. 

\noindent (b) $d = 10$ mod 8

\noindent 
The supercharges are Majorana-Weyl spinors with positive chirality 
$Q_+^i$ ($i = \! 1, 2, \cdots,$ $N_+$) and Majorana-Weyl 
spinors with negative chirality $Q_-^i$ ($i = 1, 2, \cdots, N_-$). 
The automorphism group is K = SO($N_+$) $\times$ SO($N_-$). 
Anticommutators of the supercharges are 
\ba
\{Q_+^i, Q_+^{jT} \} \A = \A {1 \over 2} 
(1+\bar\gamma) \gamma^a C_- P_a \delta^{ij}, \nonu
\{Q_-^i, Q_-^{jT} \} \A = \A {1 \over 2} 
(1-\bar\gamma) \gamma^a C_- P_a \delta^{ij}, \nonu
\{Q_+^i, Q_-^{jT} \} \A = \A {1 \over 2} 
(1+\bar\gamma) C_- Z^{ij}. 
\ea

\noindent (c) $d = 6$ mod 8

\noindent 
The supercharges are symplectic Majorana-Weyl spinors with positive 
chirality $Q_+^i$ ($i = 1, 2, \cdots, N_+$) and symplectic 
Majorana-Weyl spinors with negative chirality $Q_-^i$ 
($i = 1, 2, \cdots, N_-$). They satisfy 
$\Omega_+^{ij} (Q_+^j)^c = Q_+^i$, 
$\Omega_-^{ij} (Q_-^j)^c = Q_-^i$, 
where $\Omega_\pm^{ij}$ are antisymmetric matrices. 
The numbers $N_+$ and $N_-$ must be even. 
The automorphism group is K = USp($N_+$) $\times$ USp($N_-$). 
Anticommutators of the supercharges are 
\ba
\{Q_+^i, Q_+^{jT} \} \A = \A {1 \over 2} 
(1+\bar\gamma) \gamma^a C_- P_a \Omega_+^{ij}, \nonu
\{Q_-^i, Q_-^{jT} \} \A = \A {1 \over 2} 
(1-\bar\gamma) \gamma^a C_- P_a \Omega_-^{ij}, \nonu
\{Q_+^i, Q_-^{jT} \} \A = \A {1 \over 2} 
(1+\bar\gamma) C_- Z^{ij}. 
\ea

\noindent (d) $d = 9, 11$ mod 8 

\noindent 
The supercharges are (pseudo) Majorana spinors $Q^i$ 
($i = 1, 2, \cdots, N$). The automorphism group is 
K = SO($N$) and anticommutators of the supercharges are 
\be
\{Q^i, Q^{jT} \} = \gamma^a C P_a \delta^{ij} + C Z^{ij}, 
\ee
where $C = C_+$, $Z^{ij} = Z^{ji}$ for $d=9$ mod 8 and 
$C = C_-$, $Z^{ij} = -Z^{ji}$ for $d=11$ mod 8. 

\noindent (e) $d = 5, 7$ mod 8

\noindent 
The supercharges are symplectic (pseudo) Majorana spinors 
$Q^i$ ($i = 1, 2, \cdots, N$). 
They satisfy $\Omega^{ij} (Q^j)^c = Q^i$, where 
$\Omega^{ij}$ is an antisymmetric matrix. 
The number $N$ must be even. The automorphism group is 
K = USp($N$) and anticommutators of the supercharges are 
\be
\{Q^i, Q^{jT} \} = \gamma^a C P_a \Omega^{ij} + C Z^{ij}, 
\ee
where $C = C_+$, $Z^{ij} = - Z^{ji}$ for $d=5$ mod 8 
and $C = C_-$, $Z^{ij} = Z^{ji}$ for $d=7$ mod 8. 
\par
Particles appearing in supergravities belong to irreducible 
representations of these super Poincar\'e algebras. 
All states in an irreducible representation are obtained by 
applying components of the supercharges with helicity 
${1 \over 2}$ $Q_{1 \over 2}$ on the lowest helicity state 
$\ket{h_{\rm min}}$. A state $Q_{1 \over 2} Q_{1 \over 2} 
\cdots Q_{1 \over 2} \ket{h_{\rm min}}$ with $n$ 
$Q_{1 \over 2}$'s has helicity $h = h_{\rm min} + {1 \over 2} n$. 
(For details, see ref.\ \cite{STRATHDEE}.) 
When the supercharges have too many components, 
all representations contain particles with helicity $|h| > 2$. 
However, consistent interacting theories with helicity $> 2$ are not 
known. Hence, we should consider algebras which have representations 
with helicity $\leq 2$. Then, there are only a finite number of 
possible $(d, N)$. 
In particular, the space-time dimension must be $d \leq 11$. 
Supermultiplets for these $(d, N)$ were given in 
ref.\ \cite{STRATHDEE}. 
Representations which contain graviton and gravitinos are 
called supergravity multiplets. They are massless 
representations of the algebra, which satisfy 
\be
P_a P^a = 0, \qquad Z^{ij} = 0. 
\ee
Field contents corresponding to supergravity multiplets in 
various dimensions are listed in Table \ref{tablethree}. 
In addition to these supergravity multiplets there can exist 
massless and massive matter supermultiplets containing particles 
with spins $\leq 1$. We do not discuss matter supermultiplets 
in this paper. 
\begin{table}[tbp]
\begin{center}
\begin{tabular}{|r|c|c|l|r|} \hline
\multicolumn{1}{|c|}{$d$} &
\multicolumn{1}{|c|}{$N$} &
\multicolumn{1}{|c|}{spinors} &
\multicolumn{1}{|c|}{fields} &
\multicolumn{1}{|c|}{$n$} \\ \hline
%
11 & 1 & M & 
$e_\mu{}^a$, $\psi_\mu$, $B_{\mu\nu\rho}$ 
& 128 \\ \hline
%
%
10 & (1,1) & MW & 
$e_\mu{}^a$, $\psi_{+\mu}$, $\psi_{-\mu}$, $B_{\mu\nu\rho}$, 
$B_{\mu\nu}$, $B_\mu$, $\lambda_+$, $\lambda_-$, $\phi$ 
& 128 \\ \cline{2-5}
& (2,0) & MW & 
$e_\mu{}^a$, $2\, \psi_{+\mu}$, $B^{(+)}_{\mu\nu\rho\sigma}$, 
$2\, B_{\mu\nu}$, $2\, \lambda_-$, $2\, \phi$ 
& 128 \\ \cline{2-5}
& (1,0) & MW & 
$e_\mu{}^a$, $\psi_{+\mu}$, $B_{\mu\nu}$, $\lambda_-$, $\phi$ 
& 64 \\ \hline
%
%
9 & 2 & pM & 
$e_\mu{}^a$, $2\, \psi_\mu$, $B_{\mu\nu\rho}$, 
$2\, B_{\mu\nu}$, $3\, B_\mu$, $4\, \lambda$, $3\, \phi$ 
& 128 \\ \cline{2-5}
& 1 & pM & 
$e_\mu{}^a$, $\psi_\mu$, $B_{\mu\nu}$, $B_\mu$, $\lambda$, $\phi$ 
& 56 \\ \hline
%
%
8 & 2 & pM & 
$e_\mu{}^a$, $2\, \psi_\mu$, $B_{\mu\nu\rho}$, 
$3\, B_{\mu\nu}$, $6\, B_\mu$, $6\, \lambda$, $7\, \phi$ 
& 128 \\ \cline{2-5}
& 1 & pM & 
$e_\mu{}^a$, $\psi_\mu$, $B_{\mu\nu}$, $2\, B_\mu$, $\lambda$, $\phi$ 
& 48 \\ \hline
%
%
7 & 4 & sM & 
$e_\mu{}^a$, $4\, \psi_\mu$, $5\, B_{\mu\nu}$, $10\, B_\mu$, 
$16\, \lambda$, $14\, \phi$ 
& 128 \\ \cline{2-5}
& 2 & sM & 
$e_\mu{}^a$, $2\, \psi_\mu$, $B_{\mu\nu}$, $3\, B_\mu$, 
$2\, \lambda$, $\phi$ 
& 40 \\ \hline
%
%
6 & (4,4) & sMW & 
$e_\mu{}^a$, $4\, \psi_{+\mu}$, $4\, \psi_{-\mu}$, 
$5\, B_{\mu\nu}$, $16\, B_\mu$, $20\, \lambda_+$, 
$20\, \lambda_-$, $25\, \phi$ 
& 128 \\ \cline{2-5}
& (4,2) & sMW & 
$e_\mu{}^a$, $4\, \psi_{+\mu}$, $2\, \psi_{-\mu}$, 
$5\, B^{(+)}_{\mu\nu}$, $B^{(-)}_{\mu\nu}$, $8\, B_\mu$, 
$10\, \lambda_+$, $4\, \lambda_-$, $5\, \phi$ 
& 64 \\ \cline{2-5}
& (2,2) & sMW & 
$e_\mu{}^a$, $2\, \psi_{+\mu}$, $2\, \psi_{-\mu}$, $B_{\mu\nu}$, 
$4\, B_\mu$, $2\, \lambda_+$, $2\, \lambda_-$, $\phi$ 
& 32 \\ \cline{2-5}
& (4,0) & sMW & 
$e_\mu{}^a$, $4\, \psi_{+\mu}$, $5\, B^{(+)}_{\mu\nu}$ 
& 24 \\ \cline{2-5}
& (2,0) & sMW & 
$e_\mu{}^a$, $2\, \psi_{+\mu}$, $B^{(+)}_{\mu\nu}$ 
& 12 \\ \hline
%
%
5 & 8 & spM & 
$e_\mu{}^a$, $8\, \psi_\mu$, $27\, B_\mu$, $48\, \lambda$, $42\, \phi$ 
& 128 \\ \cline{2-5}
& 6 & spM & 
$e_\mu{}^a$, $6\, \psi_\mu$, $15\, B_\mu$, $20\, \lambda$, $14\, \phi$ 
& 64 \\ \cline{2-5}
& 4 & spM & 
$e_\mu{}^a$, $4\, \psi_\mu$, $6\, B_\mu$, $4\, \lambda$, $\phi$ 
& 24 \\ \cline{2-5}
& 2 & spM & 
$e_\mu{}^a$, $2\, \psi_\mu$, $B_\mu$ 
& 8 \\ \hline
%
%
4 & 8 & M & 
$e_\mu{}^a$, $8\, \psi_\mu$, $28\, B_\mu$, $56\, \lambda$, $70\, \phi$ 
& 128 \\ \cline{2-5}
& 6 & M & 
$e_\mu{}^a$, $6\, \psi_\mu$, $16\, B_\mu$, $26\, \lambda$, $30\, \phi$ 
& 64 \\ \cline{2-5}
& 5 & M & 
$e_\mu{}^a$, $5\, \psi_\mu$, $10\, B_\mu$, $11\, \lambda$, $10\, \phi$ 
& 32 \\ \cline{2-5}
& 4 & M & 
$e_\mu{}^a$, $4\, \psi_\mu$, $6\, B_\mu$, $4\, \lambda$, $2\, \phi$ 
& 16 \\ \cline{2-5}
& 3 & M & 
$e_\mu{}^a$, $3\, \psi_\mu$, $3\, B_\mu$, $\lambda$ 
& 8 \\ \cline{2-5}
& 2 & M & 
$e_\mu{}^a$, $2\, \psi_\mu$, $B_\mu$ 
& 4 \\ \cline{2-5}
& 1 & M & 
$e_\mu{}^a$, $\psi_\mu$ 
& 2 \\ \hline
%
\end{tabular}
\end{center}
\caption{Supergravity multiplets. 
$e_\mu{}^a$, $\psi_\mu$, $B_{\mu \cdots}$, $\lambda$ and $\phi$ 
represent vielbein, Rarita-Schwinger fields, antisymmetric tensor 
fields, spin ${1 \over 2}$ spinor fields and scalar fields 
respectively. The subscripts $\pm$ on spinor fields denote 
chiralities. The superscripts $(\pm)$ on antisymmetric tensor 
fields mean that they are (anti-)self-dual. 
The numbers of fields are counted by real fields for bosonic 
fields and (symplectic/pseudo) Majorana(-Weyl) spinors for 
fermionic fields. The last column $n$ denotes bosonic 
(= fermionic) physical degrees of freedom. 
} 
\label{tablethree}
\end{table}
\par
As a check of the field contents in Table \ref{tablethree} 
one can count the numbers of bosonic and fermionic degrees of 
freedom in a supergravity multiplet, which should be the same. 
The physical degrees of freedom of each fields are most easily 
obtained in the light-cone gauge, where only transverse 
components of the fields are physical. We find that 
the numbers of physical degrees of freedom are 
\ba
e_\mu{}^a & : & {\textstyle{1 \over 2}} \, (d-2)(d-1) - 1, \nonu
B_{\mu\nu\rho\sigma} & : & {}_{d-2}{\rm C}_4 
= {\textstyle{1 \over 24}} \, (d-2)(d-3)(d-4)(d-5), \nonu
B_{\mu\nu\rho} & : & {}_{d-2}{\rm C}_3 
= {\textstyle{1 \over 6}} \, (d-2)(d-3)(d-4), \nonu
B_{\mu\nu} & : & {}_{d-2}{\rm C}_2 
= {\textstyle{1 \over 2}} \, (d-2)(d-3), \nonu
B_\mu & : & d-2, \nonu
\phi & : & 1, \nonu
\psi_\mu & : & {\textstyle{1 \over 2}} \, (d-2-1) \, 
2^{\left[{d \over 2}\right]}, \nonu
\lambda & : & {\textstyle{1 \over 2}} \, 2^{\left[{d \over 2}\right]}. 
\ea
The number $-1$ for $e_\mu{}^a$ and $\psi_\mu$ comes from 
the ($\gamma$-)traceless conditions on graviton and gravitino. 
The factor ${1 \over 2}$ for spinor fields is due to the fact that 
their field equations are first order differential equations. 
The numbers of the bosonic and fermionic degrees of freedom 
in each supergravity multiplet are indeed the same 
and are given in the last column of the table. 
%
%
\newsection{Supergravities in higher dimensions}
Lagrangians and local supertransformation laws of fields of 
supergravities in various dimensions were explicitly 
obtained by the Noether's method or by dimensional 
reductions from higher dimensional theories. 
Supergravity in the highest space-time dimensions is the 
$d=11$, $N=1$ theory \cite{CJS}. 
The field content is the vielbein 
$e_\mu{}^a$, a Majorana Rarita-Schwinger field $\psi_\mu$ and 
a real third rank antisymmetric tensor field $B_{\mu\nu\rho}$. 
The Lagrangian has a relatively simple form 
\ba
{\cal L} \A = \A - {1 \over 4} e R 
- {1 \over 2} i e \bar\psi_\mu \gamma^{\mu\nu\rho} D_\nu \psi_\rho 
- {1 \over 48} e F_{\mu\nu\rho\sigma} F^{\mu\nu\rho\sigma} \nonu
\A \A + {1 \over 96} e \left( 
\bar\psi_\mu \gamma^{\mu\nu\alpha\beta\gamma\delta} \psi_\nu 
+ 12 \, \bar\psi^\alpha \gamma^{\beta\gamma} \psi^\delta \right)
F_{\alpha\beta\gamma\delta} \nonu
\A \A + {2 \over 144^2} \, 
\epsilon^{\alpha_1 \cdots \alpha_4 \beta_1 
\cdots \beta_4 \mu\nu\rho} F_{\alpha_1 \cdots \alpha_4} 
F_{\beta_1 \cdots \beta_4} B_{\mu\nu\rho} \ 
+ \ (\mbox{4-fermi terms}), 
\ea
where $F_{\mu\nu\rho\sigma} = 4 \partial_{[\mu} B_{\nu\rho\sigma]}$ 
is the field strength of the antisymmetric tensor. 
This Lagrangian is invariant under the local supertransformation 
\ba
\delta_Q e_\mu{}^a \A = \A 
- i \bar\epsilon \gamma^a \psi_\mu, \qquad
\delta_Q B_{\mu\nu\rho} = {3 \over 2} \, 
\bar\epsilon \gamma_{[\mu\nu} \psi_{\rho]}, \nonu
\delta_Q \psi_\mu \A = \A D_\mu \epsilon 
+ {i \over 144} \left( \gamma^{\alpha\beta\gamma\delta}{}_\mu 
- 8 \gamma^{\beta\gamma\delta} \delta_\mu^\alpha \right) 
\epsilon F_{\alpha\beta\gamma\delta} 
+ (\mbox{3-fermi terms}) 
\ea
in addition to the general coordinate and the local Lorentz 
transformations. 
It is also invariant under the local gauge transformation of 
the antisymmetric tensor field 
\be
\delta_g e_\mu{}^a = 0, \qquad 
\delta_g \psi_\mu = 0, \qquad 
\delta_g B_{\mu\nu\rho} 
= 3 \, \partial_{[\mu} \Lambda_{\nu\rho]} \qquad 
(\Lambda_{\mu\nu} = - \Lambda_{\nu\mu}). 
\ee
\par
In ten dimensions there are three types of supergravities: 
$(N_+, N_-) = (1,1)$, $(2,0)$, $(1,0)$. 
The field contents are given in Table \ref{tablethree}. 
The $(1,1)$ supergravity \cite{GP} can be obtained from 
the $d=11$ theory by a dimensional reduction and is vector-like 
(left-right symmetric). This theory is a massless sector of 
the type IIA superstring theory. 
The $(2,0)$ supergravity \cite{SHW} is a chiral 
(left-right asymmetric) theory. 
It contains a fourth rank antisymmetric tensor field 
$B_{\mu\nu\rho\sigma}^{(+)}$, whose field strength satisfies 
a self-duality condition 
$F_{\mu\nu\rho\sigma\tau} = {1 \over 5!} e 
\epsilon_{\mu\nu\rho\sigma\tau\alpha\beta\gamma\delta\eta} 
F^{\alpha\beta\gamma\delta\eta}$. 
Because of this self-dual field a Lorentz covariant action of 
this theory is not known although field equations were 
explicitly obtained. It has a non-compact symmetry SU(1,1) and 
the scalar fields are described by an SU(1,1)/U(1) non-linear 
sigma model. This theory is a massless sector of the type IIB 
superstring theory. 
The $(1,0)$ supergravity \cite{Bddv} is a chiral theory. 
There exists a matter supermultiplet $(A_\mu, \chi_+)$, where 
$A_\mu$ is a gauge field and $\chi_+$ is a spin ${1 \over 2}$ 
Majorana-Weyl spinor field, both of which are in the adjoint 
representation of a certain gauge group. 
This theory is a massless sector of the type I superstring theory 
and the heterotic string theory. 
\par
Supergravities in $d<10$ dimensions, whose field contents 
are given in Table \ref{tablethree}, can be obtained from 
$d=11$ or $d=10$ supergravity by dimensional reductions and 
truncations of fields. Their general structure 
is as follows. Field contents are the vielbein, Rarita-Schwinger 
fields, antisymmetric tensor gauge fields, spin ${1 \over 2}$ 
spinor fields and scalar fields. When scalar fields are present, 
the theory has a rigid non-compact symmetry G. The scalar fields 
are described by a G/H non-linear sigma model, where H is a maximal 
compact subgroup of G. For instance, two scalar fields in the 
$d=4$, $N=4$ theory are described by the SU(1,1)/U(1) sigma model. 
In even dimensions ($d = 2n$) G acts 
on ($n \! - \! 1$)-th antisymmetric tensor fields 
$B_{\mu_1 \cdots \mu_{n-1}}$ as duality transformations. 
In this case G is a symmetry of equations of motion but 
not of the action. In the following two sections we will 
discuss these structures in detail. 
\newpage
%
\newsection{Non-linear sigma models}
Scalar fields appearing in supergravities are described by a 
G/H non-linear sigma model, where G is a non-compact Lie group 
and H is a maximal compact subgroup of G. 
The G/H non-linear sigma model is a theory of G/H-valued scalar 
fields, which is invariant under rigid G transformations. 
In this section we shall review how to construct G/H non-linear 
sigma models \cite{CWZ}, \cite{GZ}. 
\par
We represent the scalar fields by a G-valued scalar field $V(x)$ 
and require local H invariance. Since we do not introduce independent 
H gauge fields, the H part of $V(x)$ can be gauged away and physical 
degrees of freedom are on a coset space G/H. 
The rigid G transformations act on $V(x)$ from the left 
\be
V(x) \rightarrow g V(x) \qquad (g \in G), 
\label{gtrans}
\ee
while the local H transformations act from the right 
\be
V(x) \rightarrow V(x) h^{-1}(x) \qquad (h(x) \in H). 
\label{htrans}
\ee
\par
To construct the action we decompose the Lie algebra ${\bf G}$ 
of G as 
\be
{\bf G} = {\bf H} + {\bf N}, 
\ee
where ${\bf H}$ is the Lie algebra of H and ${\bf N}$ is its 
orthogonal complement in ${\bf G}$. 
The orthogonality is defined with respective to trace 
in a certain representation: $\tr ({\bf H} \, {\bf N}) = 0$. 
It can be easily shown that 
\be
[ {\bf H}, {\bf H} ] \subset {\bf H}, \qquad 
[ {\bf H}, {\bf N} ] \subset {\bf N}. 
\label{orthocom}
\ee
The ${\bf G}$-valued field $V^{-1} \partial_\mu V$ is 
decomposed as 
\be
V^{-1} \partial_\mu V = Q_\mu + P_\mu, \qquad 
Q_\mu \in {\bf H}, \quad P_\mu \in {\bf N}. 
\label{pqdecom}
\ee
By using eq.\ (\ref{orthocom}) the transformation laws of 
$Q_\mu$ and $P_\mu$ under the local H transformations (\ref{htrans}) 
are found to be 
\ba
Q_\mu \A \rightarrow \A h Q_\mu h^{-1} + h \partial_\mu h^{-1}, \nonu
P_\mu \A \rightarrow \A h P_\mu h^{-1}, 
\ea
while they are invariant under the rigid G 
transformations (\ref{gtrans}). We see that $Q_\mu$ transforms 
as an H gauge field, while $P_\mu$ is covariant under 
the H transformations. From eq.\ (\ref{pqdecom}) $P_\mu$ can 
be expressed as 
\be
P_\mu = V^{-1} \left( \partial_\mu V - V Q_\mu \right) 
\equiv V^{-1} D_\mu V, 
\ee
where $D_\mu$ is the H-covariant derivative on $V$. 
\par
By using these quantities we can construct an action which is 
invariant under the rigid G and the local H transformations. 
The kinetic term of the scalar fields is 
\ba
{\cal L} \A = \A {1 \over 2} \tr ( P_\mu P^\mu ) \nonu
\A = \A {1 \over 2} \tr ( V^{-1} D_\mu V V^{-1} D^\mu V ). 
\label{nsmlag}
\ea
This action is quadratic in derivatives of $V$ and is manifestly 
invariant under the rigid G and the local H transformations. 
The H-connection $Q_\mu$ can be used to define the covariant 
derivatives on other fields transforming under the local H. 
For instance, when a spinor field $\psi(x)$ transforms under 
the local H transformations as 
$\psi(x) \rightarrow h(x) \psi(x)$, 
the covariant derivative is 
\be
D_\mu \psi = \left( \partial_\mu + Q_\mu \right) \psi. 
\ee
We can also use $P_\mu$ to construct H-invariant terms 
in the action such as 
\be
\bar\psi \gamma^\mu P_\mu \psi. 
\ee
\par
We can describe the theories in terms of physical fields by fixing 
a gauge for the local H symmetry. For instance, we can choose a gauge 
\be
V(x) = \e^{\Phi(x)}, 
\ee
where $\Phi(x)$ is an {\bf N}-valued field, which represents 
physical degrees of freedom. The G transformations (\ref{gtrans}) 
break this gauge condition. 
To preserve the gauge the transformation $g$ must be 
accompanied by a compensating H transformation $h(x; g)$. 
Therefore, the G transformations of $\Phi(x)$ are given by 
\be
\e^{\Phi(x)} \rightarrow 
\e^{\Phi'(x)} = g \e^{\Phi(x)} h^{-1}(x; g). 
\ee
The compensating transformation $h(x; g)$ is chosen such that 
$\Phi'(x)$ belongs to {\bf N}. The transformation 
$\Phi(x) \rightarrow \Phi'(x)$ is a non-linear realization of 
(\ref{gtrans}). When $g \in {\rm H}$, we can take $h(x; g) = g$ 
and $g$ is linearly realized. 
\par
The groups G and H appearing in supergravities are listed 
in Table \ref{tablefour}. 
\begin{table}[tbp]
\begin{center}
\begin{tabular}{|r|c|c|c|} \hline
\multicolumn{1}{|c|}{$d$} &
\multicolumn{1}{|c|}{$N$} &
\multicolumn{1}{|c|}{G} &
\multicolumn{1}{|c|}{H} \\ \hline
10 & (1,1) & GL(1,\,\R) & 1 \\ \cline{2-4}
& (2,0) & SL(2,\,\R) & SO(2) \\ \cline{2-4}
& (1,0) & G(1,\,\R) & 1 \\ \hline
9 & 2 & GL(2,\,\R) & SO(2) \\ \cline{2-4}
& 1 & GL(1,\,\R) & 1 \\ \hline
8 & 2 & SL(3,\,\R) $\times$ SL(2,\,\R) & SO(3) 
$\times$ SO(2) \\ \cline{2-4}
& 1 & GL(1,\,\R) & 1 \\ \hline
7 & 4 & SL(5,\,\R) & SO(5) \\ \cline{2-4}
& 2 & GL(1,\,\R) & 1 \\ \hline
6 & (4,4) & SO(5,5) & SO(5) $\times$ SO(5) \\ \cline{2-4}
& (4,2) & SO(5,1) & SO(5) \\ \cline{2-4}
& (2,2) & GL(1,\,\R) & 1 \\ \hline 
5 & 8 & ${\rm E}_{6(+6)}$ & USp(8) \\ \cline{2-4}
& 6 & ${\rm SU}^*(6)$ & USp(6) \\ \cline{2-4}
& 4 & USp(4) $\times$ GL(1,\,\R) & USp(4) \\ \hline 
4 & 8 & ${\rm E}_{7(+7)}$ & SU(8) \\ \cline{2-4}
& 6 & ${\rm SO}^*$(12) & U(6) \\ \cline{2-4}
& 5 & SU(5,1) & U(5) \\ \cline{2-4}
& 4 & SU(4) $\times$ SL(2,\,\R) & U(4) \\ \hline 
\end{tabular}
\end{center}
\caption{G and H in supergravities.}
\label{tablefour}
\end{table}
One can check that the dimension of the coset space G/H is 
equal to the number of scalar fields in each theory. 
As an example of G/H sigma models in supergravities 
let us consider the case G = SL(2, {\bf R}) $\sim$ SU(1,1) 
and H = SO(2) $\sim$ U(1). This sigma model appears in the 
$d=10$, $(2,0)$ and $d=4$, $N=4$ supergravities. 
The SU(1,1)-valued scalar field $V(x)$ is parametrized by two 
complex scalar fields $\phi_0(x)$, $\phi_1(x)$ as 
\be
V(x) = 
\left(
\begin{array}{cc}
\phi_0(x) & \phi_1^*(x) \\
\phi_1(x) & \phi_0^*(x) 
\end{array}
\right), \qquad 
| \phi_0 |^2 - | \phi_1 |^2 = 1. 
\ee
The rigid SU(1,1) and the local U(1) transformations are given by 
eqs.\ (\ref{gtrans}) and (\ref{htrans}) respectively with 
\be
g = 
\left(
\begin{array}{cc}
a & b^* \\
b & a^* 
\end{array}
\right), \qquad
h(x) = 
\left(
\begin{array}{cc}
\e^{i\theta(x)} & 0 \\
0 & \e^{-i\theta(x)} 
\end{array}
\right), 
\ee
where $|a|^2 - |b|^2 = 1$. 
The quantities in eq.\ (\ref{pqdecom}) are obtained as 
\ba
Q_\mu \A = \A \left( \phi_0^* \partial_\mu \phi_0 
- \phi_1^* \partial_\mu \phi_1 \right) 
\left(
\begin{array}{cc}
1 & 0 \\
0 & -1 
\end{array}
\right), \nonu
P_\mu \A = \A 
\left(
\begin{array}{cc}
0 & (\phi_0 \partial_\mu \phi_1 - \phi_1 \partial_\mu \phi_0)^* \\
\phi_0 \partial_\mu \phi_1 - \phi_1 \partial_\mu \phi_0 & 0 
\end{array}
\right). 
\ea
Therefore, the Lagrangian (\ref{nsmlag}) becomes 
\ba
{\cal L} \A = \A {1 \over 2} \tr \left( P_\mu P^\mu \right) \nonu
\A = \A | \phi_0 \partial_\mu \phi_1 
- \phi_1 \partial_\mu \phi_0 |^2 \nonu
\A = \A {\partial_\mu z \partial^\mu z^* \over 
(1-|z|^2)^2} \qquad 
\left( z \equiv \phi^*_1 (\phi^*_0)^{-1} \right). 
\label{nsmex}
\ea
The variable $z$ is U(1)-invariant and represents physical degrees 
of freedom. It transforms under SU(1,1) as 
\be
z \rightarrow {a z + b^* \over b z + a^*}. 
\ee
It can be easily seen that the Lagrangian (\ref{nsmex}) is 
invariant under this transformation. 
%
%
\newsection{Duality symmetries}
\begin{flushleft}
7.1\ \ Duality symmetry in the free Maxwell theory
\end{flushleft}
In this section we discuss duality symmetries appearing 
in supergravities in even dimensions. Duality symmetries are 
generalizations of the electric-magnetic duality in the 
Maxwell theory. Let us first consider the free Maxwell theory 
as a simple example to explain what duality symmetries are. 
The free Maxwell equations consist of the equation of motion 
and the Bianchi identity 
\be
\partial_\mu F^{\mu\nu} = 0, \qquad
\partial_\mu \tilde F^{\mu\nu} = 0, 
\label{maxwell}
\ee
where 
\be
F_{\mu\nu} = \partial_\mu A_\nu - \partial_\nu A_\mu, \qquad
\tilde F^{\mu\nu} = {1 \over 2} \epsilon^{\mu\nu\rho\sigma} 
F_{\rho\sigma}. 
\label{dualdef}
\ee
The set of equations (\ref{maxwell}) is invariant under 
rigid general linear transformations 
\be
\delta \left(
\begin{array}{c}
F^{\mu\nu} \\
\tilde F^{\mu\nu} 
\end{array}
\right) 
= 
\left(
\begin{array}{cc}
A & B \\
C & D 
\end{array}
\right)
\left(
\begin{array}{c}
F^{\mu\nu} \\
\tilde F^{\mu\nu} 
\end{array}
\right) \qquad
(A,\; B,\; C,\; D \in \R). 
\label{maxwellgl}
\ee
Such transformations, which mix $F^{\mu\nu}$ and 
$\tilde F^{\mu\nu}$ are called duality transformations. 
We have to take into account the fact that $F^{\mu\nu}$ and 
$\tilde F^{\mu\nu}$ are not independent but are related by 
the duality operation in eq.\ (\ref{dualdef}). 
By taking a dual of the upper equation of eq.\ (\ref{maxwellgl}) 
and using an identity $\tilde{\tilde F} = - F$, we obtain 
$\delta \tilde F = A \tilde F - B F$, which should coincide 
with the lower equation. Therefore, the transformation parameters 
must satisfy $D = A$ and $C = -B$, and the symmetry 
transformation is 
\be
\delta \left(
\begin{array}{c}
F \\
\tilde F 
\end{array}
\right) 
= 
\left(
\begin{array}{cc}
A & B \\
-B & A 
\end{array}
\right)
\left(
\begin{array}{c}
F \\
\tilde F 
\end{array}
\right). 
\label{maxwellduality}
\ee
We can diagonalize the transformation matrix by using 
a complex basis 
\be
\delta \left(
\begin{array}{c}
F + i \tilde F \\
F - i \tilde F 
\end{array}
\right) 
= 
\left(
\begin{array}{cc}
A-iB & 0 \\
0 & A+iB 
\end{array}
\right)
\left(
\begin{array}{c}
F + i \tilde F \\
F - i \tilde F 
\end{array}
\right). 
\ee
We see that the group of the duality transformations 
is GL(1,\,\C). 
\par
To discuss the duality symmetry we have not studied the 
invariance of the action but that of the field equations. 
The reason is that the duality transformations are consistent 
only on-shell. Since the independent variables of the theory 
is $A_\mu$, $\delta F_{\mu\nu}$ in eq.\ (\ref{maxwellduality}) 
should be derived from $\delta A_\mu$: 
\be
\partial_\mu \delta A_\nu - \partial_\nu \delta A_\mu 
= A F_{\mu\nu} + B \tilde F_{\mu\nu}. 
\ee
The integrability of this equation requires 
$\partial_\mu (A \tilde F + B \tilde{\tilde F})^{\mu\nu} = 0$, 
i.e., $\partial_\mu F^{\mu\nu} = 0$. 
Thus, the equation of motion must be satisfied. 
Even if we ignore this point and formally consider the 
transformation (\ref{maxwellduality}) off-shell, the action 
$-{\textstyle{1 \over 4}} \int d^4 x F_{\mu\nu}^2$ is not 
invariant. Therefore, to construct theories invariant under 
duality transformations it is easier to study the covariance 
of equations of motion. 
\par
\begin{flushleft}
7.2\ \ Duality symmetries in higher dimensions 
\end{flushleft}
We shall study duality symmetries of interacting theories in 
general even dimensions $d = 2n$ \cite{CJ}, \cite{GZ}, 
\cite{TANII}. We consider theories of 
$(n-1)$-th rank antisymmetric tensor fields 
$B^I_{\mu_1 \cdots \mu_{n-1}}(x)$ ($I = 1, \cdots, M$) 
interacting with other fields $\phi_i(x)$. 
The field strengths of the tensor fields and their duals 
are defined as 
\ba
F^I_{\mu_1 \cdots \mu_n} 
\A = \A n \, \partial_{[\mu_1} B^I_{\mu_2 \cdots \mu_n]}, \nonu
\tilde F^{I \mu_1 \cdots \mu_n} 
\A = \A {1 \over n!} e^{-1} 
\epsilon^{\mu_1 \cdots \mu_n \nu_1 \cdots \nu_n} 
F^I_{\nu_1 \cdots \nu_n}. 
\ea
In $d$ dimensions the duality operation satisfies 
\be
\tilde{\tilde F} = \epsilon F, \qquad 
\epsilon = \cases{+1 & for $d=4k+2$, \cr
                  -1 & for $d=4k$. \cr
}
\ee
We assume that the Lagrangian has a form 
\ba
{\cal L} \A = \A {\cal L}(\phi, \partial\phi, F) \nonu 
\A = \A {1 \over 2n!} \epsilon e K_{1IJ}(\phi) 
F^I_{\mu_1 \cdots \mu_n} F^{J \mu_1 \cdots \mu_n} 
+ {1 \over 2n!} \epsilon e K_{2IJ}(\phi) F^I_{\mu_1 \cdots \mu_n} 
\tilde F^{J \mu_1 \cdots \mu_n} \nonu 
\A \A + e F^I_{\mu_1 \cdots \mu_n} 
O_I^{\mu_1 \cdots \mu_n}(\phi, \partial\phi) 
+ {\cal L}'(\phi, \partial\phi), 
\ea
where $K_{1IJ} = K_{1JI}$, $K_{2IJ} = - \epsilon K_{2JI}$. 
The fields $B^I_{\mu_1 \cdots \mu_{n-1}}$ appear only 
through their field strengths $F^I_{\mu_1 \cdots \mu_n}$. 
The Lagrangians of supergravities are of this type. 
We require duality symmetries in this theory, and 
obtain conditions on the functions $K_1$, $K_2$, $O$ 
and possible duality symmetry groups. 
\par
The equations of motion for $B^I_{\mu_1 \cdots \mu_{n-1}}$ 
and the Bianchi identities are 
\be
\partial_{\mu_1} \left( e \tilde G_I^{\mu_1 \cdots \mu_n} 
\right) = 0, \qquad
\partial_{\mu_1} \left( e \tilde F^{I \, \mu_1 \cdots \mu_n} 
\right) = 0, 
\ee
where the dual of antisymmetric tensors 
$G_{I \mu_1 \cdots \mu_n}$ are defined by 
\be
\tilde G_I^{\mu_1 \cdots \mu_n} 
= {n! \over e} {\partial {\cal L} \over \partial 
F^I_{\mu_1 \cdots \mu_n}}. 
\label{gdef}
\ee
(For the free Maxwell theory, $G^{\mu\nu} = \tilde F^{\mu\nu}$.) 
These equations are invariant under transformations 
\be
\delta \left(
\begin{array}{c}
F \\
G 
\end{array}
\right) 
= 
\left(
\begin{array}{cc}
A & B \\
C & D 
\end{array}
\right)
\left(
\begin{array}{c}
F \\
G 
\end{array}
\right), \qquad \delta \phi^i = \xi^i(\phi), 
\label{gltrans}
\ee
where $A$, $B$, $C$, $D$ are constant $n \times n$ real matrices 
and $\xi^i(\phi)$ are functions of $\phi^i$. 
As in the Maxwell theory these constants are not independent. 
We shall obtain the conditions that these constants should 
satisfy by studying (i) the covariance of the definition of $G$ 
(\ref{gdef}) and (ii) the covariance of the equations of motion 
for $\phi^i$. 
\par
Let us first study the covariance of the definition of $G$. 
By eq.\ (\ref{gdef}) $G$ is expressed in terms of 
$F$ and $\phi$. Therefore, the transformation of $G$ can be 
derived from those of $F$ and $\phi$. From eq.\ (\ref{gdef}) 
we obtain 
\be
\delta \tilde G_I 
= {n! \over e} {\partial \delta {\cal L} \over \partial F^I} 
- \tilde G_J A^J{}_I 
- \tilde G_J B^{JK} {\partial G_K \over \partial F^I}. 
\ee
This should coincide with the transformation given in the 
lower equations in eq.\ (\ref{gltrans}). 
By equating these two transformation laws we obtain 
\ba
\A \A {\partial \over \partial F^I} \left( 
n! \delta {\cal L} - {1 \over 2} e F^J C_{JK} \tilde F^K 
- {1 \over 2} e \tilde G_J B^{JK} G_K \right) 
- \left( A^J{}_I + D_I{}^J \right) 
n! {\partial {\cal L} \over \partial F^J} \nonu
\A \A \qquad\qquad\qquad
= {1 \over 2} e \left( C_{IJ} + \epsilon C_{JI} \right) \tilde F^J 
+ {1 \over 2} e \tilde G_J \left( B^{JK} + \epsilon B^{KJ} \right) 
{\partial G_K \over \partial F^I}. 
\ea
When there exist nontrivial interactions, this equation gives 
conditions on the transformation parameters 
\be
A^I{}_J + D_J{}^I = \alpha \delta^I_J, \qquad
B^{IJ} = - \epsilon B^{JI}, \qquad
C_{IJ} = - \epsilon C_{JI}, 
\label{conditionone}
\ee
where $\alpha$ is an arbitrary constant, as well as a condition 
on the variation of the Lagrangian 
\be
{\partial \over \partial F^I} \left(
\delta {\cal L} - {1 \over 2n!} e F^I C_{IJ} \tilde F^J 
- {1 \over 2n!} e \tilde G_I B^{IJ} G_J 
- \alpha {\cal L} \right) = 0. 
\label{conditiontwo}
\ee
The equation of motion for $\phi^i$ is 
\be
E_i \equiv 
\left(
{\partial \over \partial \phi^i} - \partial_\mu 
{\partial \over \partial (\partial_\mu \phi^i)} \right)
{\cal L} = 0. 
\ee
The covariance of this equation under the duality transformations 
(\ref{gltrans}) 
\be
\delta E_i = - {\partial \xi^j \over \partial \phi^i} E_j 
\ee
requires another condition on the variation of the Lagrangian 
\be
\left(
{\partial \over \partial \phi^i} - \partial_\mu 
{\partial \over \partial (\partial_\mu \phi^i)} \right) 
\left( \delta {\cal L} - {1 \over 2n!} e \tilde G_I B^{IJ} G_J 
\right) = 0. 
\label{conditionthree}
\ee
\par
Now we can find out possible duality groups by studying 
eqs.\ (\ref{conditionone}), (\ref{conditiontwo}) and 
(\ref{conditionthree}). Comparing eqs.\ (\ref{conditiontwo}) 
and (\ref{conditionthree}) we find $\alpha = 0$. 
Then, the conditions on the parameters (\ref{conditionone}) can 
be written as 
\be
X^T \Omega + \Omega X = 0, 
\ee
where 
\be
X = 
\left(
\begin{array}{cc}
A & B \\
C & D 
\end{array}
\right), 
\qquad
\Omega = 
\left(
\begin{array}{cc}
0 & \epsilon \\
1 & 0 
\end{array}
\right). 
\ee
For $d = 4k$ ($\epsilon = -1$) $\Omega$ is an antisymmetric matrix 
and the above condition implies that the group of duality 
transformations is Sp($2M$,\,\R) or its subgroup. 
On the other hand, for $d = 4k+2$ ($\epsilon = +1$) $\Omega$ is a 
symmetric matrix, which can be diagonalized 
to diag$({\bf 1}, -{\bf 1})$. Therefore, the group of duality 
transformations in this case is SO($M$,\,$M$) or its subgroup. 
Eqs.\ (\ref{conditionone}), (\ref{conditiontwo}) and 
(\ref{conditionthree}) also restrict the variation of the Lagrangian 
\ba
\delta {\cal L} \A = \A {1 \over 2n!} e F^I C_{IJ} \tilde F^J 
+ {1 \over 2n!} e \tilde G_I B^{IJ} G_J \nonu
\A = \A \delta \left( {1 \over 2n!} e F^I \tilde G_I \right). 
\label{ltransf}
\ea
Thus, although the Lagrangian is not invariant under the duality 
transformations, it transforms in a definite way. 
\par
It can be shown that a derivative of the Lagrangian with 
respect to an invariant parameter $\lambda$ is invariant 
under the duality transformations. 
Indeed, by computing 
${\partial \over \partial\lambda} \delta {\cal L}$ and 
using eqs.\ (\ref{gdef}), (\ref{conditionone}) we obtain 
\be
\delta \left( {\partial {\cal L} \over \partial \lambda} \right) 
= {\partial \over \partial\lambda} \left( 
\delta {\cal L} - {1 \over 2n!} e F^I C_{IJ} \tilde F^J 
- {1 \over 2n!} e \tilde G_I B^{IJ} G_J 
\right), 
\ee
which vanishes by eq.\ (\ref{ltransf}). 
Here, we have assumed that $\xi^i$ do not depend on $\lambda$. 
The invariant parameter can be an invariant external field 
such as the metric. Thus, the energy-momentum tensor 
obtained as a functional derivative of the Lagrangian with 
respect to the metric is invariant under the duality 
transformations. 
\par
Let us obtain an explicit form of the Lagrangian which 
transforms as in eq.\ (\ref{ltransf}). 
The Lagrangian satisfying eq.\ (\ref{ltransf}) can be written as 
\ba
{\cal L} \A = \A {1 \over 2n!} e F^I \tilde G_I 
+ {\cal L}_{\rm inv}(\phi, \partial\phi, F) \nonu
\A = \A {1 \over 2n!} e F^I \tilde G_I 
+ {1 \over 2n!} e ( F^I I_I + \epsilon G_I H^I ) 
+ {\cal L}_{\rm inv}(\phi, \partial\phi), 
\label{duallag}
\ea
where $n$-th antisymmetric tensors 
$( H^I_{\mu_1 \cdots \mu_n}(\phi, \partial\phi), 
I_{I \,\mu_1 \cdots \mu_n}(\phi, \partial\phi) )$ 
transform in the same way as $(F^I, G_I)$, and 
${\cal L}_{\rm inv}(\phi, \partial\phi)$ is invariant under 
the duality transformations. 
In the second line we have assumed that the duality symmetry 
group is a maximal one, i.e., Sp($2M$,\,\R) in $d=4k$ or 
SO($M$,\,$M$) in $d=4k+2$, for simplicity. 
When the symmetry group is a subgroup 
of them, there can be other invariants other than 
$F I + \epsilon G H$. 
Substituting eq.\ (\ref{duallag}) into eq.\ (\ref{gdef}) 
we obtain a differential equation for $\tilde G$ 
\be
(\tilde G - I)_I = (F - \epsilon \tilde H)^J 
{\partial \over \partial F^I} (\tilde G - I)_J. 
\ee
To solve this equation we introduce an operation $j$: 
\be
j :\ \ F\  \longrightarrow\  jF \equiv \tilde F. 
\ee
Then, the solution can be written as 
\be
j G_I = I_I + \epsilon K_{IJ}(\phi) ( F^J - \epsilon j H^J ), 
\label{gsolution}
\ee
where 
\be
K_{IJ}(\phi) = K_{1IJ}(\phi) + K_{2IJ}(\phi) j \quad
\left( K_{1IJ} = K_{1JI}, \ \ 
K_{2IJ} = - \epsilon K_{2JI} \right). 
\ee
{}From the covariance of eq.\ (\ref{gsolution}) under 
the duality transformations, $K$ must transform as 
\be
\delta K = - K A - K B K j + \epsilon C j + D K. 
\label{ktransf}
\ee
Substituting this solution into eq.\ (\ref{duallag}) we obtain 
\be
{\cal L} = {1 \over 2n!} \epsilon e F^I K_{IJ} F^J 
+ {1 \over n!} e F^I ( I_I - K_{IJ} \tilde H^J) 
- {1 \over 2n!} \epsilon e \tilde H^I ( I_I - K_{IJ} \tilde H^J) 
+ {\cal L}_{\rm inv}(\phi, \partial\phi). 
\label{duallagrangian}
\ee
Thus, if we can find out functions $H^I(\phi, \partial\phi)$, 
$I^I(\phi, \partial\phi)$, $K_{IJ}(\phi)$ with appropriate 
transformation properties, we have an explicit form 
of the  Lagrangian. 
\par
\begin{flushleft}
7.3\ \ Compact duality symmetry 
\end{flushleft}
Let us consider a special case of $K = 1$. 
In this case we will see that the duality symmetry group 
must be a compact group. From eq.\ (\ref{ktransf}) 
the transformation parameters must satisfy 
\be
A = D, \qquad B = \epsilon C. 
\label{compactcond}
\ee
\par
For $d = 4k$ these conditions imply 
\be
X = 
\left(
\begin{array}{cc}
A & B \\
-B & A 
\end{array}
\right), \qquad
A^T = -A, \quad B^T = B. 
\ee
The transformation law becomes 
\be
\delta \left(
\begin{array}{c}
F + i G \\
F - i G 
\end{array}
\right) 
= 
\left(
\begin{array}{cc}
A-iB & 0 \\
0 & (A-iB)^* 
\end{array}
\right)
\left(
\begin{array}{c}
F + i G \\
F - i G 
\end{array}
\right). 
\ee
Since $A-iB$ is anti-hermitian, the duality symmetry group is 
U($M$), which is a maximal compact subgroup of Sp($M$,\,$M$), 
or its subgroup. 
\par
On the other hand, for $d = 4k +2$ the conditions in 
eq.\ (\ref{compactcond}) imply 
\be
X = 
\left(
\begin{array}{cc}
A & B \\
B & A 
\end{array}
\right), \qquad
A^T = -A, \quad B^T = -B. 
\ee
and 
\be
\delta \left(
\begin{array}{c}
F + G \\
F - G 
\end{array}
\right) 
= 
\left(
\begin{array}{cc}
A+B & 0 \\
0 & A-B 
\end{array}
\right)
\left(
\begin{array}{c}
F + G \\
F - G 
\end{array}
\right). 
\ee
Since $A+B$ and $A-B$ are real and antisymmetric, the duality 
symmetry group is SO($M$) $\times$ SO($M$), which is 
a maximal compact subgroup of SO($M$,\,$M$), or its subgroup. 
\par
As an example of compact duality symmetries let us consider 
the $d=4$, $N=2$ supergravity \cite{FSZ}. 
The fields are the vierbein $e_\mu{}^a$, two Majorana 
Rarita-Schwinger fields $\psi_\mu^i$ ($i=1,2$) and 
a U(1) gauge field $B_\mu$. The Lagrangian is 
\ba
{\cal L} \A = \A - {1 \over 4} e R - {1 \over 2} i e 
\bar\psi_\mu^i \gamma^{\mu\nu\rho} D_\nu \psi_\rho^i 
- {1 \over 4} e F_{\mu\nu} F^{\mu\nu} \nonu
\A \A - {1 \over 2} e \epsilon_{ij} \bar\psi_\mu^i 
( F^{\mu\nu} - i \bar\gamma \tilde F^{\mu\nu} ) \psi_\nu^j 
+ (\mbox{4-fermi terms}). 
\label{sugracompactlag}
\ea
This Lagrangian is invariant under a rigid SU(2) transformation 
\be
\delta e_\mu{}^a = 0, \qquad 
\delta \psi_\mu^i = ( \Sigma^{ij} 
+ i \bar\gamma \Lambda^{ij} ) \psi_\mu^j, \qquad
\delta B_\mu = 0, 
\ee
where $\Sigma^{ij}$, $\Lambda^{ij}$ are are real parameters 
satisfying $\Sigma^{ij} = - \Sigma^{ji}$, 
$\Lambda^{ij} = \Lambda^{ji}$ and $\Lambda^{ii} = 0$. 
The equations of motion have an additional 
symmetry under a rigid U(1) transformation 
\be
\delta e_\mu{}^a = 0, \qquad 
\delta \psi_\mu^i 
= -{1 \over 2} \Lambda i \bar\gamma \psi_\mu^i, \qquad
\delta \left(
\begin{array}{c}
F^{\mu\nu} \\
G^{\mu\nu} 
\end{array}
\right) 
= 
\left(
\begin{array}{cc}
0 & \Lambda \\
- \Lambda  & 0 
\end{array}
\right)
\left(
\begin{array}{c}
F^{\mu\nu} \\
G^{\mu\nu} 
\end{array}
\right), 
\label{sugracompact}
\ee
where $\Lambda$ is a real parameter. 
This U(1) transformation acts on the gauge field as a duality 
transformation. An explicit form of $G$ can be obtained from 
eq.\ (\ref{gdef}) 
\be
\tilde G_{\mu\nu} = - F_{\mu\nu} - \tilde H_{\mu\nu} + I_{\mu\nu}, 
\label{compactg}
\ee
where 
\be
H_{\mu\nu} = - \epsilon_{ij} \bar\psi_\mu^i i \bar\gamma 
\psi_\nu^j, \qquad 
I_{\mu\nu} = - \epsilon_{ij} \bar\psi_\mu^i \psi_\nu^j. 
\ee
One can easily see that $(H, I)$ transform in the same way as 
$(F, G)$. Furthermore, the transformation of $G$ in 
eq.\ (\ref{compactg}) derived from $\delta F$ and $\delta \psi$ 
correctly reproduces $\delta G$ in eq.\ (\ref{sugracompact}). 
Using $G$ the Lagrangian (\ref{sugracompactlag}) can be 
rewritten in the form (\ref{duallag}) 
\be
{\cal L} = {1 \over 4} e F_{\mu\nu} \tilde G^{\mu\nu} 
+ {1 \over 4} e \left( F_{\mu\nu} I^{\mu\nu} 
- G_{\mu\nu} H^{\mu\nu} \right)
+ (\mbox{$B_\mu$-independent terms}). 
\ee
\par
\begin{flushleft}
7.4\ \ Non-compact duality symmetry 
\end{flushleft}
We can construct $K_{IJ}(\phi)$ which transforms as in 
eq.\ (\ref{ktransf}) by using a G/H non-linear sigma model. 
Here, G is a duality symmetry group, which we assume to be a 
maximal one, i.e., Sp($2M$) or SO($M$,\,$M$). H is a maximal 
compact subgroup of G, i.e., U($M$) or SO($M$) $\times$ SO($M$). 
We use a G-valued scalar field $V(x)$, which transforms under 
${\rm G}_{\rm rigid} \times {\rm H}_{\rm local}$ as 
in eqs.\ (\ref{gtrans}), (\ref{htrans}). 
\par
Let us first discuss the case $d=4k$, G = Sp($2M$), H = U($M$). 
It is convenient to use the complex basis, 
in which the G transformation is 
\be
\left(
\begin{array}{c}
F + i G \\
F - i G 
\end{array}
\right) 
\rightarrow 
\left(
\begin{array}{cc}
a & b^* \\
b & a^* 
\end{array}
\right) 
\left(
\begin{array}{c}
F + i G \\
F - i G 
\end{array}
\right), 
\qquad
a^\dagger a - b^\dagger b = 1, \quad 
a^T b - b^T a = 0. 
\ee
In this basis the scalar field is expressed as 
\be
V(x) = 
\left(
\begin{array}{cc}
\phi_0(x) & \phi_1^*(x) \\
\phi_1(x) & \phi_0^*(x)
\end{array}
\right), 
\qquad
\phi_0^\dagger \phi_0 - \phi_1^\dagger \phi_1 = 1, \quad 
\phi_0^T \phi_1 - \phi_1^T \phi_0 = 0. 
\label{4kscalar}
\ee
Using the components in eq.\ (\ref{4kscalar}) 
we can construct $K$ transforming as in eq.\ (\ref{ktransf})
\be
K = (\phi_0^* - \phi_1^*) (\phi_0^* + \phi_1^*)^{-1}, 
\ee
where the imaginary unit $i$ in $\phi_0$, $\phi_1$ is replaced 
by the operation $j$, i.e., 
$\phi_0 = {\rm Re}\, \phi_0 + j\, {\rm Im}\, \phi_0$, 
$\phi_0^* = {\rm Re}\, \phi_0 - j\, {\rm Im}\, \phi_0$, etc. 
Note that $j^2 = -1$ in $d=4k$ as $i^2 = -1$. 
It is convenient to introduce an H-invariant variable 
\be
z = \phi_1^* (\phi_0^*)^{-1} = z^T, 
\ee
which transforms under G as 
\be
z \rightarrow (a z + b^*) (b z + a^*)^{-1}. 
\ee
Then, $K$ can be expressed as 
\be
K = {1-z \over 1+z}. 
\ee
\par
We now turn to $d = 4k+2$ and consider the maximal case 
G = SO($M$,$M$), H = SO($M$) $\times$ SO($M$). In the 
$\Omega$-diagonal basis the G transformation is written as 
\ba
\A\A \left(
\begin{array}{c}
F + G \\
F - G 
\end{array}
\right) 
\rightarrow 
\left(
\begin{array}{cc}
a & b \\
c & d 
\end{array}
\right) 
\left(
\begin{array}{c}
F + G \\
F - G 
\end{array}
\right), \nonu
\A\A 
a^T a - c^T c = 1, \qquad 
d^T d - b^T b = 1, \qquad 
a^T b - c^T d = 0. 
\ea
The G-valued scalar field is parametrized as 
\ba
\A\A V(x) = 
\left(
\begin{array}{cc}
\phi_1(x) & \psi_2(x) \\
\psi_1(x) & \phi_2(x)
\end{array}
\right), \nonu
\A\A 
\phi_1^T \phi_1 - \psi_1^T \psi_1 = 1, \qquad 
\phi_2^T \phi_2 - \psi_2^T \psi_2 = 1, \qquad 
\phi_1^T \psi_2 - \psi_1^T \phi_2 = 0. 
\ea
In this case the operation $j$ satisfies $j^2 = 1$, which suggests 
to introduce the projection operators 
\be
P_\pm = {1 \over 2} (1 \pm j). 
\ee
A coefficient function $K$ which has the right transformation 
property in eq. (\ref{ktransf}) is 
\be
K = (\phi_1 - \psi_1) (\phi_1 + \psi_1)^{-1} P_+ 
+ (\phi_2 - \psi_2) (\phi_2 + \psi_2)^{-1} P_-. 
\ee
We define an H-invariant variable 
\be
z = ( \psi_1 (\phi_1)^{-1} )^T = \psi_2 (\phi_2)^{-1}, 
\ee
which transforms under the G transformation as 
\be
z \rightarrow (az+b)(cz+d)^{-1}. 
\ee
In terms of this variable $K$ can be written as 
\be
K = {1-z^T \over 1+z^T} P_+ + {1-z \over 1+z} P_-. 
\ee
\par
As an example of non-compact duality symmetries in supergravities 
let us consider the $d=4$, $N=4$ theory \cite{CSF}, \cite{CJ}. 
Among the fields in the theory we are interested in six U(1) gauge 
fields $B^{ij}_\mu = - B^{ji}_\mu$ ($i, j = 1, \cdots, 4$) and 
two real scalar fields (see Table \ref{tablethree}). 
The scalar fields are represented as the G/H non-linear sigma 
model with G = SU(4) $\times$ SU(1,1), H = SU(4) $\times$ U(1). 
The SU(4) factors cancel each other in the coset G/H. 
G = SU(4) $\times$ SU(1,1) acts on the gauge fields as duality 
transformations. The SU(4) transformation is 
\be
\left(
\begin{array}{c}
(F + iG)^{ij} \\
(F - iG)^{ij} 
\end{array}
\right) 
\rightarrow 
\left(
\begin{array}{cc}
U^{ik} U^{jl} & 0 \\
0 & U^{*ik} U^{*jl} 
\end{array}
\right) 
\left(
\begin{array}{c}
(F + iG)^{kl} \\
(F - iG)^{kl} 
\end{array}
\right), 
\ee
where $U^\dagger U = 1$, $\det U = 1$, while the SU(1,1) 
transformation is 
\be
\left(
\begin{array}{c}
(F + iG)^{ij} \\
(F - iG)^{ij} 
\end{array}
\right) 
\rightarrow 
\left(
\begin{array}{cc}
a \delta^{i[k} \delta^{l]j} & b^* {1 \over 2} \epsilon^{ijkl} \\
b {1 \over 2} \epsilon^{ijkl} & a^* \delta^{i[k} \delta^{l]j} 
\end{array}
\right) 
\left(
\begin{array}{c}
(F + iG)^{kl} \\
(F - iG)^{kl} 
\end{array}
\right), 
\ee
where $a, b \in \C$, $|a|^2 - |b|^2 = 1$. 
The SU(1,1)-valued scalar field is parametrized as 
\be
V(x) = 
\left(
\begin{array}{cc}
\Phi_0(x) & \Phi_1^*(x) \\
\Phi_1(x) & \Phi^*_0(x) 
\end{array}
\right), 
\ee
where 
\ba
& & \Phi_0^{ijkl}(x) 
= \phi_0(x) \delta^{i[k} \delta^{l]j}, \qquad 
\Phi_1^{ijkl}(x) 
= \phi_1(x) {1 \over 2} \epsilon^{ijkl}. \nonu
&& \qquad \phi_0(x), \phi_1(x) \in \C, \quad 
|\phi_0|^2 - |\phi_1|^2 =1. 
\ea
The imaginary unit $i$ in $\phi_0$ and $\phi_1$ is replaced by 
the duality operation $j$. In terms of an H-invariant variable 
\ba
Z^{ij, kl}(x) \A = \A (\Phi^*_1(x))^{ij, pq} 
(\Phi^*_0(x)^{-1})^{pq,kl} \nonu
\A = \A z(x) {1 \over 2} \epsilon^{ijkl}, 
\ea
where $z(x) = \phi^*_1(x) \phi^*_0(x)^{-1}$ 
we construct 
\ba
K_{ij, kl} \A = \A \left( {1-Z \over 1+Z} \right)^{ij, kl} \nonu 
\A = \A {1+z^2 \over 1-z^2} \delta^{i[k} \delta^{l]j} 
- {2z \over 1-z^2} {1 \over 2} \epsilon^{ijkl}. 
\ea
Then, the Lagrangian can be written as 
\be
{\cal L} 
= e {\partial_\mu z \partial^\mu z^* \over (1-|z|^2)^2} 
- {1 \over 4} e F_{\mu\nu}^{ij} K_{ij, kl} F^{\mu\nu kl} 
+ \cdots. 
\ee
%
%
\newsection{Super{\boldmath $p$}-branes}
In this section we briefly discuss super $p$-branes, 
which are closely related to supergravities.  
We shall consider a theory of $p$-dimensionally extended 
objects moving in $d$-dimensional space-time. 
They are generalizations of strings and are called $p$-branes: 
0-branes are point particles, 1-branes are strings, 
2-branes are membranes, etc. 
\par
Let us first consider bosonic $p$-branes without supersymmetry. 
Dynamical variables are $X^\mu(\xi)$ ($\mu = 0, 1, \cdots, d-1$), 
which represent a map from $(p+1)$-dimensional world-volume 
with coordinates $\xi^i$ ($i = 0, 1, \cdots, p$) to 
$d$-dimensional space-time. 
When space-time is a flat Minkowski space-time, a natural action 
is the one proportional to the volume of the $(p+1)$-dimensional 
world volume (Nambu-Goto type action) 
\be
S[X] = - T \int d^{p+1} \xi \sqrt{ | \det h_{ij} | }, \qquad
h_{ij} = \partial_i X^\mu \partial_j X^\nu \eta_{\mu\nu}, 
\label{ngaction}
\ee
where $T$ is the $p$-brane tension 
of dimension $({\rm length})^{-p-1}$. 
We will take $T=1$ in the following for simplicity. 
$h_{ij}$ is a metric on the world-volume induced by 
the space-time flat metric $\eta_{\mu\nu}$. 
This action is invariant under the space-time Poincar\'e 
transformations and the world-volume reparametrizations. 
One can write down another action (Polyakov type action) 
using an independent metric $g_{ij}$ 
\be
S'[X,g] = - \int d^{p+1} \xi \left[ 
{1 \over 2} \sqrt{ | g | } g^{ij} 
\partial_i X^{\mu} \partial_j X^{\nu} \eta_{\mu \nu} 
- {1 \over 2} (p-1) \sqrt{ | g | } \right], 
\label{paction}
\ee
which are equivalent to the above action (\ref{ngaction}). 
The field equation of $g_{ij}$ can be solved 
algebraically: $g_{ij} = h_{ij}$. 
(For $p=1$ the general solution is $g_{ij} = \e^{\phi} h_{ij}$, 
where $\phi$ is an arbitrary function of $\xi^i$.) 
Substituting this solution into eq.\ (\ref{paction}) we obtain 
the action (\ref{ngaction}). Therefore, these two actions are 
equivalent, at least at the classical level. 
We will use the Nambu-Goto type action (\ref{ngaction}) 
in the following. 
\par
We now consider a supersymmetric generalization of $p$-branes. 
In the case of strings ($p=1$) there are two formulations: 
the Neveu-Schwarz-Ramond formulation and the Green-Schwarz 
formulation. For general $p$-branes Neveu-Schwarz-Ramond 
formulation is not known. A technical reason is that 
a supersymmetrization of the cosmological term 
$\sqrt{ | g | }$ for $p > 1$ requires the Einstein 
term $\sqrt{ | g | }R$ and the theory becomes more complicated. 
Moreover, even if one could construct an appropriate action with 
world-volume local supersymmetry, it is not clear whether 
it leads to space-time supersymmetry. 
On the other hand, the Green-Schwarz formulation of 
$p$-branes was constructed in refs.\ \cite{HLP}, 
\cite{AETW}, which we will discuss in the following. 
\par
Dynamical variables of super $p$-branes in the Green-Schwarz 
formulation are $Z^M(\xi) = (X^\mu(\xi), \theta^{\alpha}(\xi))$ 
$(M=(\mu,\alpha), \, \mu=0,1,\cdots, d-1; \, \alpha=1, \cdots, n)$, 
which represent a map from $(p+1)$-dimensional world-volume to 
$d$-dimensional (extended) superspace. 
Here, $n$ is a number of independent components of fermionic 
coordinates of the superspace. 
The action of super $p$-branes in Minkowski space-time is 
\ba
\A\A S = - \int d^{p+1} \xi \left[ \sqrt{ | \det h_{ij} | } 
+ {2 \over (p+1)!} 
\epsilon^{i_1 \cdots i_{p+1}} \Pi^{A_1}_{i_1} \cdots 
\Pi^{A_{p+1}}_{i_{p+1}} B_{A_{p+1} \cdots A_1} \right], \nonu
\A \A \qquad \Pi^{\mu}_i = \partial_iX^{\mu}-i\bar \theta 
\gamma^{\mu} \partial_i \theta, \qquad 
\Pi^{\alpha}_i = \partial_i\theta^\alpha, \qquad 
h_{ij} = \Pi^{\mu}_i \Pi^{\nu}_j \eta_{\mu\nu}, 
\label{superngaction}
\ea
where $\gamma^{\mu}$ are $d$-dimensional gamma matrices. 
$B_{A_{p+1} \cdots A_1}(Z)$ $(A=(\mu, \alpha))$ is a $(p+1)$-form 
on the superspace, whose non-vanishing components of 
the field strength $H_{A_{p+2} \cdots A_1}$ are 
\be
H_{\alpha \beta \mu_1 \cdots \mu_p} 
= - i \zeta^{-1} (C^{-1 T} 
\gamma_{\mu_1 \cdots \mu_p})_{\alpha \beta}, \qquad
\zeta = (-1)^{{1 \over 4}p(p-1)}. 
\label{hcomponent}
\ee
\par
The action (\ref{superngaction}) is invariant under the 
space-time super Poincar\'e transformations and 
$(p+1)$-dimensional reparametrizations on the world-volume. 
The space-time supertransformations are 
\be
\delta_Q X^\mu = i \bar\epsilon \gamma^\mu \theta, \qquad
\delta_Q \theta = \epsilon, 
\ee
where $\epsilon^\alpha$ is a constant spinor parameter. 
In the case of superstrings it is also invariant under local 
fermionic transformations called $\kappa$-transformations. 
This symmetry reduces the fermionic degrees of freedom by half. 
We require such symmetry also for $p \ge 2$. 
The $\kappa$-transformations are 
\ba
\A\A \delta_\kappa X^{\mu} 
= i \bar \theta \gamma^{\mu} \delta_\kappa \theta, \qquad
\delta_\kappa \theta = (1+\Gamma)\kappa, \nonu 
\A\A \Gamma \equiv {\zeta \over (p+1)! \sqrt{|h|}} 
\epsilon^{i_1 \cdots i_{p+1}} \Pi^{\mu_1}_{i_1} \cdots 
\Pi^{\mu_{p+1}}_{i_{p+1}} \gamma_{\mu_1 \cdots \mu_{p+1}}, 
\label{kappatransformation}
\ea
where $\kappa^\alpha(\xi)$ is a parameter of the transformations. 
The matrix $\Gamma$ defined above satisfies $\Gamma^2 = 1$ and 
therefore ${1 \over 2}(1 \pm \Gamma)$ are projection operators. 
The action (\ref{superngaction}) is invariant under the 
transformations (\ref{kappatransformation}) provided that 
the field strength of $B_{A_{p+1} \cdots A_1}$ is given by 
eq.\ (\ref{hcomponent}). Thus, the presence of the second term 
(Wess-Zumino term) in the action (\ref{superngaction}) is required 
by the $\kappa$-invariance. 
\par
\begin{figure}[tb]
\setlength{\unitlength}{0.9mm}
\begin{picture}(150,145)(-5,10)
%
\thicklines
\put(20,20){\vector(1,0){120}}
\put(20,20){\vector(0,1){130}}
\thinlines
\put(40,20){\line(0,1){2}}
\put(60,20){\line(0,1){2}}
\put(80,20){\line(0,1){2}}
\put(100,20){\line(0,1){2}}
\put(120,20){\line(0,1){2}}
\put(20,30){\line(1,0){2}}
\put(20,40){\line(1,0){2}}
\put(20,50){\line(1,0){2}}
\put(20,60){\line(1,0){2}}
\put(20,70){\line(1,0){2}}
\put(20,80){\line(1,0){2}}
\put(20,90){\line(1,0){2}}
\put(20,100){\line(1,0){2}}
\put(20,110){\line(1,0){2}}
\put(20,120){\line(1,0){2}}
\put(20,130){\line(1,0){2}}
\put(20,140){\line(1,0){2}}
\put(37,14){\makebox(6,6)[b]{1}}
\put(57,14){\makebox(6,6)[b]{2}}
\put(77,14){\makebox(6,6)[b]{3}}
\put(97,14){\makebox(6,6)[b]{4}}
\put(117,14){\makebox(6,6)[b]{5}}
\put(137,14){\makebox(6,6)[b]{$p$}}
\put(10,27){\makebox(7,6)[r]{1}}
\put(10,37){\makebox(7,6)[r]{2}}
\put(10,47){\makebox(7,6)[r]{3}}
\put(10,57){\makebox(7,6)[r]{4}}
\put(10,67){\makebox(7,6)[r]{5}}
\put(10,77){\makebox(7,6)[r]{6}}
\put(10,87){\makebox(7,6)[r]{7}}
\put(10,97){\makebox(7,6)[r]{8}}
\put(10,107){\makebox(7,6)[r]{9}}
\put(10,117){\makebox(7,6)[r]{10}}
\put(10,127){\makebox(7,6)[r]{11}}
\put(10,137){\makebox(7,6)[r]{12}}
\put(10,147){\makebox(7,6)[r]{$d$}}
\put(40,50){\line(2,1){20}}
\put(40,60){\line(2,1){40}}
\put(40,80){\line(2,1){80}}
\put(40,120){\line(2,1){20}}
\put(40,50){\circle*{2}}
\put(40,60){\circle*{2}}
\put(40,80){\circle*{2}}
\put(40,120){\circle*{2}}
\put(60,60){\circle*{2}}
\put(60,70){\circle*{2}}
\put(60,90){\circle*{2}}
\put(60,130){\circle*{2}}
\put(80,80){\circle*{2}}
\put(80,100){\circle*{2}}
\put(100,110){\circle*{2}}
\put(120,120){\circle*{2}}
\put(64,59){\makebox(4,6)[l]{$1=1$}}
\put(84,79){\makebox(4,6)[l]{$2=2$}}
\put(124,119){\makebox(4,6)[l]{$4=4$}}
\put(64,129){\makebox(4,6)[l]{$8=8$}}
%
%
\end{picture}
\caption{The brane scan.}
\label{figureone}
\end{figure}
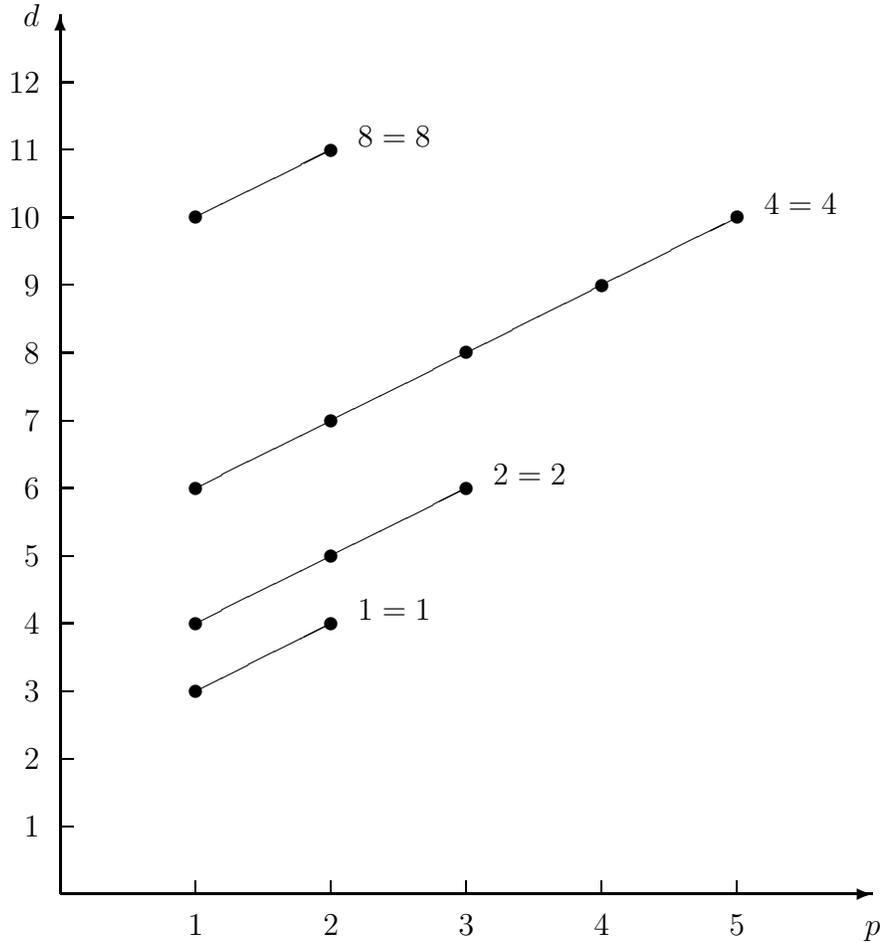

The field strength $H_{A_{p+2} \cdots A_1}$ given in 
eq.\ (\ref{hcomponent}) must be a closed $(p+2)$-form. 
This requires that gamma matrices should satisfy a certain kind 
of identity. When $\theta$ is a Majorana spinor, the identity is 
\be
(C^{-1} \gamma_{\mu_1})_{(\alpha \beta} 
(C^{-1} \gamma^{\mu_1 \cdots \mu_p})_{\gamma \delta )} = 0. 
\label{gammaid}
\ee
A similar identity must be satisfied 
when $\theta$ is a spinor of other type. 
(For details see ref.\ \cite{AETW}.) 
These identities lead to a condition on $d$, $p$ and $n$ 
\be
d-p-1 = {1 \over 4} n 
\label{bfeq}
\ee
for $p \geq 2$ and 
\be
d-2 = {1 \over 4} n \qquad \mbox{or} \qquad d-2 = {1 \over 2} n
\label{bfeq2}
\ee
for $p=1$. 
The left and the right hand sides of eqs.\ (\ref{bfeq}) and 
(\ref{bfeq2}) represent bosonic ($X^\mu$) and fermionic ($\theta$) 
physical degrees of freedom up to gauge degrees of freedom 
respectively. Eqs.\ (\ref{bfeq}) and (\ref{bfeq2}) are satisfied 
only for 12 pairs $(d,p)$ shown in Fig.\ \ref{figureone}. 
Possible types of supersymmetries are 
$(N_+,N_-)$ = (1,1), (2,0), (1,0) for $(d,p)$ = $(10,1)$, 
$(N_+,N_-)$ = (2,2), (4,0), (2,0) for $(d,p)$ = $(6,1)$, 
$N=$ 2, 1 for $(d,p)$ = (4,1), (3,1) 
and the minimal one for other $(d,p)$. 
The numbers on the right of each sequence in Fig.\ \ref{figureone} 
represent physical degrees of freedom (\ref{bfeq}), (\ref{bfeq2}). 
(A similar analysis for general signature of space-time 
was given in ref.\ \cite{BD}.) 
%
\par
The action (\ref{superngaction}) is the one for a flat Minkowski 
space-time. One can introduce space-time background fields. 
The background fields are represented by supervielbein $E_M{}^A$ 
and super $(p+1)$-form $B_{A_{p+1} \cdots A_1}$, both of which 
are superfields. The action is given by eq.\ (\ref{superngaction}) 
with $\Pi_i^A$ replaced by 
\be
\Pi_i^A = \partial_i Z^M E_M{}^A. 
\ee
The $\kappa$-invariance of the action imposes 
constraints on these background fields. 
For $(d,p) = (11,2)$, $(10,1)$, $(10,5)$ these constraints are shown 
to be equivalent to field equations of $d=11$, $N=1$ and $d=10$ (1,1) 
supergravities in superspace. 
For other values of $(d,p)$ an equivalence to field equations of 
supergravities is expected but has not yet been analyzed in detail. 
For string theories $(p=1)$ there is one-to-one correspondence 
between background fields and massless physical states. 
It is an interesting open problem whether such relations 
are present for $p \geq 2$. 
\par
\vspace{5mm}
%
%
The author would like to thank T. Eguchi and N. Sakai for 
suggesting him to publish this work. 
\vspace{5mm} 
%
\def\numberbysectiona{\@addtoreset{equation}{section}
\def\theequation{A.\arabic{equation}}}
\numberbysectiona
\vspace{7mm}
\pagebreak[3]
\setcounter{section}{1}
\setcounter{equation}{0}
\setcounter{subsection}{0}
\setcounter{footnote}{0}
\begin{center}
{\large{\bf Appendix A. The vielbein formalism}}
\end{center}
\nopagebreak
\medskip
\nopagebreak
\hspace{3mm}
In the usual formulation of gravitational theories the 
metric tensor $g_{\mu\nu}(x)$ ($\mu, \nu = 0, 1, \cdots, d-1$) 
is used to describe gravity. 
Our signature convention of $g_{\mu\nu}$ is $(+, -, \cdots, -)$. 
The Einstein Lagrangian in this formulation is 
\be
{\cal L} = -{1 \over 16\pi G} \sqrt{-g} R, 
\label{a1}
\ee
where $G$ is the gravitational constant
and $g = \det g_{\mu\nu}$. In the following 
we will put $4\pi G = 1$ for simplicity. 
The scalar curvature $R$ is defined from the Ricci 
tensor $R_{\mu\nu}$ and the Riemann tensor 
$R_{\mu\nu}{}^\rho{}_\sigma$ as 
\ba
\A\A R = g^{\mu\nu} R_{\mu\nu}, \qquad 
R_{\mu\nu} = R_{\rho\mu}{}^\rho{}_\nu, \nonu
\A\A R_{\mu\nu}{}^\rho{}_\sigma 
= \partial_\mu \Gamma_{\nu\sigma}^{\,\rho} 
- \partial_\nu \Gamma_{\mu\sigma}^{\,\rho} 
+ \Gamma_{\mu\lambda}^{\,\rho} \Gamma_{\nu\sigma}^{\,\lambda} 
- \Gamma_{\nu\lambda}^{\,\rho} \Gamma_{\mu\sigma}^{\,\lambda}. 
\label{a2}
\ea
The Christoffel symbol $\Gamma_{\mu\nu}^{\,\lambda}$ is defined 
by using the metric as 
\be
\Gamma_{\mu\nu}^{\,\lambda} = {1 \over 2} g^{\lambda\rho} 
\left( 
\partial_\mu g_{\nu\rho} + \partial_\nu g_{\mu\rho} 
- \partial_\rho g_{\mu\nu} \right). 
\label{a3}
\ee
This form is determined by the metricity condition 
$D_\lambda g_{\mu\nu} \equiv \partial_\lambda g_{\mu\nu} 
- \Gamma_{\lambda\mu}^{\,\rho} g_{\rho\nu} 
- \Gamma_{\lambda\nu}^{\,\rho} g_{\mu\rho} = 0$ 
and the torsionless condition 
$\Gamma_{\mu\nu}^{\,\lambda} = \Gamma_{\nu\mu}^{\,\lambda}$. 
\par
To couple gravity to spinor fields it is more convenient to 
use the vielbein formulation of gravity. 
In this formulation we introduce $d$ vectors 
$e_a{}^\mu(x)$ ($a = 0, 1, \cdots, d-1$) 
at each point of space-time, 
which are orthogonal to each other and have a unit length 
\be
e_a{}^\mu(x) e_b{}^\nu(x) g_{\mu\nu}(x) = \eta_{ab}, 
\label{a4}
\ee
where $\eta_{ab} = {\rm diag}(+1, -1, \cdots, -1)$ is 
a flat Minkowski metric. We also introduce inverse matrices 
$e_\mu{}^a(x)$, which satisfy 
\be
e_\mu{}^a(x) e_a{}^\nu(x) = \delta_\mu^\nu, \qquad 
e_a{}^\mu(x) e_\mu{}^b(x) = \delta_a^b. 
\label{a5}
\ee
The fields $e_\mu{}^a(x)$ are called vielbein (vierbein or 
tetrad in four dimensions, f\"unfbein in five dimensions, 
etc.). From eqs.\ (\ref{a4}) and (\ref{a5}) we can express 
the metric in terms of the vielbein 
\be
g_{\mu\nu}(x) = e_\mu{}^a(x) e_\nu{}^b(x) \eta_{ab}. 
\label{a6}
\ee
Therefore, we can use the vielbein $e_\mu{}^a(x)$ as 
dynamical variables representing gravitational degrees of freedom. 
\par
For a given metric $g_{\mu\nu}$ the vielbein $e_\mu{}^a$ 
satisfying eq.\ (\ref{a6}) is not uniquely determined. 
If $e_\mu{}^a$ satisfies eq.\ (\ref{a6}), then another vielbein 
\be
e'_\mu{}^a(x) = e_\mu{}^b(x) \Lambda_b{}^a(x) \qquad 
\left( \Lambda_a{}^c(x) \Lambda_b{}^d(x) \eta_{cd} = \eta_{ab} 
\right)
\label{a7}
\ee
also satisfies eq.\ (\ref{a6}) with the same $g_{\mu\nu}$. 
The metric tensor has ${1 \over 2} d (d+1)$ independent 
components, while the vielbein has $d^2$ components. 
The difference ${1 \over 2} d (d-1)$ is the number of 
independent components of $\Lambda_a{}^b$. 
The transformation (\ref{a7}) is called a local Lorentz 
transformation. Since the theories are originally formulated 
by using only $g_{\mu\nu}$, they should be invariant under 
the local Lorentz transformations. 
Thus, gravitational theories in the vielbein formulation 
have two local symmetries: the general coordinate symmetry 
and the local Lorentz symmetry. 
\par
We have now two kinds of vector indices: $\mu, \nu, \cdots$ and 
$a, b, \cdots$. To distinguish them the indices 
$\mu, \nu, \cdots$ are called `world indices', while 
$a, b, \cdots$ are called `local Lorentz indices'. 
These two kinds of indices are converted into each other 
by using the vielbein and its inverse, e.g., 
\be
A_a(x) = e_a{}^\mu(x) A_\mu(x), \qquad 
A_\mu(x) = e_\mu{}^a(x) A_a(x). 
\label{a8}
\ee
Tensor fields with local Lorentz indices transform under 
the local Lorentz transformations as in eq.\ (\ref{a7}). 
They also transform under the general coordinate transformations 
as tensor fields determined by the world indices they have. 
For instance, the general coordinate ($G$) and the local 
Lorentz ($L$) transformations of a tensor field $T_{\mu a}(x)$ are 
\ba
\delta_G T_{\mu a} \A = \A - \xi^\nu \partial_\nu T_{\mu a} 
- \partial_\mu \xi^\nu T_{\nu a}, \nonu
\delta_L T_{\mu a} \A = \A - \lambda_a{}^b T_{\mu b}, 
\label{a9}
\ea
where $\xi(x)$ and $\lambda^{ab}(x) = - \lambda^{ba}(x)$ 
are infinitesimal transformation parameters. 
The transformations of spinor fields are assumed to be 
\ba
\delta_G \psi \A = \A - \xi^\mu \partial_\mu \psi, \nonu
\delta_L \psi 
\A = \A - {1 \over 4} \lambda_{ab} \gamma^{ab} \psi. 
\label{a10}
\ea
Under the general coordinate transformations they transform as 
scalars. 
\par
To construct an action of spinor fields invariant under 
the local Lorentz transformations we need a gauge field. 
It is called a spin connection $\omega_\mu{}^a{}_b(x)$ 
($\omega_\mu{}^{ab} = - \omega_\mu{}^{ba}$). 
The local Lorentz transformation of the spin connection 
should be 
\be
\delta_L \omega_\mu{}^{ab} 
= D_\mu \lambda^{ab} 
\equiv \partial_\mu \lambda^{ab} 
+ \omega_\mu{}^a{}_c \lambda^{cb} 
+ \omega_\mu{}^b{}_c \lambda^{ac} 
\label{a11}
\ee
so that the covariant derivative of a spinor field $\psi$ 
\be
D_\mu \psi 
= \left( \partial_\mu + {1 \over 4} \omega_\mu{}^{ab} 
\gamma_{ab} \right) \psi 
\label{a12}
\ee
transforms covariantly. The spinor Lagrangian invariant under 
the general coordinate and the local Lorentz transformations is 
\be
{\cal L} = i e \, \bar\psi \gamma^\mu D_\mu \psi, 
\ee
where $e = {\rm det}\, e_\mu{}^a = \sqrt{-g}$ and 
$\gamma^\mu = \gamma^a e_a{}^\mu$. 
\par
As for the Christoffel symbol, the spin connection is completely 
determined by the vielbein if we impose the torsionless condition 
\be
D_\mu e_\nu{}^a - D_\nu e_\mu{}^a = 0 \qquad
(D_\mu e_\nu{}^a 
\equiv \partial_\mu e_\nu{}^a + \omega_\mu{}^a{}_b e_\nu{}^b). 
\label{a13}
\ee
(The metricity condition corresponds to the antisymmetry property 
$\omega_\mu{}^{ab} = - \omega_\mu{}^{ba}$.) 
The solution of eq.\ (\ref{a13}) is 
$\omega_{\mu ab} = \omega_{\mu ab}(e)$, where 
\be
\omega_{\mu ab}(e) = {1 \over 2} \left( 
e_a{}^\nu \Omega_{\mu\nu b} - e_b{}^\nu \Omega_{\mu\nu a} 
- e_a{}^\rho e_b{}^\sigma e_\mu{}^c \Omega_{\rho\sigma c} 
\right), \quad
\Omega_{\mu\nu a} = \partial_\mu e_{\nu a} - \partial_\nu e_{\mu a}. 
\label{a14}
\ee
The spin connection (\ref{a14}) is related to the Christoffel 
symbol (\ref{a3}) as 
\be
\partial_\mu e_\nu{}^a + \omega_\mu{}^a{}_b e_\mu{}^b 
- \Gamma_{\mu\nu}^{\,\lambda} e_\lambda{}^a = 0. 
\label{a15}
\ee
The field strength of the spin connection 
\be
R_{\mu\nu}{}^a{}_b 
= \partial_\mu \omega_\nu{}^a{}_b 
- \partial_\nu \omega_\mu{}^a{}_b 
+ \omega_\mu{}^a{}_c \omega_\nu{}^c{}_b 
- \omega_\nu{}^a{}_c \omega_\mu{}^c{}_b. 
\label{a16}
\ee
is shown to be related to the Riemann tensor and 
the scalar curvature as 
\be
R_{\mu\nu}{}^\rho{}_\sigma 
= R_{\mu\nu}{}^a{}_b e_a{}^\rho e_\sigma{}^b, \qquad
R = e_a{}^\mu e_b{}^\nu R_{\mu\nu}{}^{ab}. 
\label{a17}
\ee
\par
%
%
\def\numberbysectionb{\@addtoreset{equation}{section}
\def\theequation{B.\arabic{equation}}}
\numberbysectionb
\vspace{7mm}
\pagebreak[3]
\setcounter{section}{1}
\setcounter{equation}{0}
\setcounter{subsection}{0}
\setcounter{footnote}{0}
\begin{center}
{\large{\bf Appendix B. Local supersymmetry invariance 
of {\boldmath $d=$} 4, {\boldmath $N=$} 1 supergravity}}
\end{center}
\nopagebreak
\medskip
\nopagebreak
\hspace{3mm}
The Lagrangian of $d=4$, $N=1$ supergravity consists of two terms 
\ba
{\cal L} \A = \A {\cal L}_{\rm E} + {\cal L}_{\rm RS}, \nonu
{\cal L}_{\rm E} 
\A = \A - {1 \over 4} e \, e_a{}^\mu e_b{}^\nu 
\hat R_{\mu\nu}{}^{ab} 
= {1 \over 16} \epsilon^{\mu\nu\rho\sigma} \epsilon_{abcd} 
e_\rho{}^c e_\sigma{}^d \hat R_{\mu\nu}{}^{ab}, \nonu
{\cal L}_{\rm RS} \A = \A - {1 \over 2} i e e_a{}^\mu e_b{}^\nu 
e_c{}^\rho \bar\psi_\mu \gamma^{abc} \hat D_\nu \psi_\rho 
= {1 \over 2} \epsilon^{\mu\nu\rho\sigma} \bar\psi_\mu 
\gamma_\nu \gamma_5 \hat D_\rho \psi_\sigma, 
\label{b1}
\ea
where $\epsilon^{\mu\nu\rho\sigma}$ is the totally antisymmetric 
tensor with $\epsilon^{0123} = +1$ and 
$\gamma_5 = i \gamma^0 \gamma^1 \gamma^2 \gamma^3$. 
The Riemann tensor $\hat R_{\mu\nu}{}^{ab}$ and the covariant 
derivative $\hat D_\mu$ depend on the vierbein $e_\mu{}^a$ only 
through the spin connection $\hat\omega_\mu{}^{ab}$. 
When the action is viewed as a functional of $e_\mu{}^a$, 
$\psi_\mu$ and $\hat\omega_\mu{}^{ab}$, the spin connection 
(\ref{sct}) satisfies an equation 
\be
{\delta  \over \delta \hat\omega_{\mu ab}} 
\int d^4 x \, {\cal L}(e, \psi, \hat\omega)= 0. 
\label{eqofomega}
\ee
(To show this it is convenient to use the second form of 
${\cal L}_{\rm E}$ and ${\cal L}_{\rm RS}$ in eq.\ (\ref{b1}).) 
Therefore, when we compute a variation of the Lagrangian under 
supertransformations, the spin connection need not be varied. 
\par
To show the local supersymmetry invariance we need the following 
formulae involving spinors. For four arbitrary spinors $\psi$, 
$\chi$, $\lambda$ and $\phi$ the Fierz identity 
\be
\bar\psi \chi \bar\lambda \phi 
= -{1 \over 4} \left[ 
\bar\psi \phi \bar\lambda \chi 
+ \bar\psi \gamma^a \phi \bar\lambda \gamma_a \chi 
- {1 \over 2} \bar\psi \gamma^{ab} \phi 
\bar\lambda \gamma_{ab} \chi 
- \bar\psi \gamma^a \gamma_5 \phi 
\bar\lambda \gamma_a \gamma_5 \chi 
+ \bar\psi \gamma_5 \phi \bar\lambda \gamma_5 \chi 
\right]
\label{fierzid}
\ee
is satisfied. Bilinears of two arbitrary Majorana spinors 
$\psi$ and $\chi$ have symmetry properties 
\ba
\bar\psi \chi \A = \A \bar\chi \psi, \nonu
\bar\psi \gamma^a \chi \A = \A - \bar\chi \gamma^a \psi, \nonu
\bar\psi \gamma^{ab} \chi \A = \A - \bar\chi \gamma^{ab} \psi, \nonu
\bar\psi \gamma^a \gamma_5 \chi 
\A = \A \bar\chi \gamma^a \gamma_5 \psi, \nonu
\bar\psi \gamma_5 \chi \A = \A \bar\chi \gamma_5 \psi. 
\label{bilinearid}
\ea
\par
Let us now compute the variation of the Lagrangian (\ref{b1}) 
under the supertransformation (\ref{localsusy}). 
Using the first form in eq.\ (\ref{b1}) the variation of 
the Einstein term is 
\ba
\delta_Q {\cal L}_{\rm E} 
\A = \A - {1 \over 4} \delta_Q (e \, e_a{}^\mu e_b{}^\nu) 
\hat R_{\mu\nu}{}^{ab} \nonu
\A = \A - {1 \over 2} i e \bar\epsilon \gamma^\mu \psi_a 
\left( e_b{}^\nu \hat R_{\mu\nu}{}^{ab} 
- {1 \over 2} e_\mu{}^a \hat R \right). 
\label{evar}
\ea
On the other hand, using the second form in eq.\ (\ref{b1}) 
the variation of the Rarita-Schwinger term is 
\ba
\delta_Q {\cal L}_{\rm RS} 
\A = \A {1 \over 2} \epsilon^{\mu\nu\rho\sigma} 
\delta_Q \bar\psi_\mu \gamma_\nu \gamma_5 \hat D_\rho \psi_\sigma 
+ {1 \over 2} \epsilon^{\mu\nu\rho\sigma} \bar\psi_\mu 
\gamma_\nu \gamma_5 \hat D_\rho \delta_Q \psi_\sigma \nonu
\A \A + {1 \over 2} \epsilon^{\mu\nu\rho\sigma} 
\delta_Q e_\nu{}^a \; 
\bar\psi_\mu \gamma_a \gamma_5 \hat D_\rho \psi_\sigma. 
\label{rsvar}
\ea
By partial integration the first term becomes 
\ba
{1 \over 2} \epsilon^{\mu\nu\rho\sigma} \hat D_\mu 
\bar\epsilon \gamma_\nu \gamma_5 \hat D_\rho \psi_\sigma 
\A = \A - {1 \over 2} \epsilon^{\mu\nu\rho\sigma} \bar\epsilon 
\gamma_\nu \gamma_5 \hat D_\mu \hat D_\rho \psi_\sigma 
- {1 \over 2} \epsilon^{\mu\nu\rho\sigma} \hat D_\mu e_\nu{}^a \, 
\bar\epsilon \gamma_a \gamma_5 \hat D_\rho \psi_\sigma \nonu
\A \A + \mbox{ total derivative terms}. 
\label{rsvar2}
\ea
By using eq.\ (\ref{torsion}), the Fierz identity 
(\ref{fierzid}) and the symmetry properties (\ref{bilinearid}) 
the second term in eq.\ (\ref{rsvar2}) is shown to cancel the 
third term in eq.\ (\ref{rsvar}). Then, eq.\ (\ref{rsvar}) becomes 
\ba
\delta_Q {\cal L}_{\rm RS} 
\A = \A - {1 \over 4} \epsilon^{\mu\nu\rho\sigma} 
\bar\epsilon \gamma_\nu \gamma_5 [ \hat D_\mu, \hat D_\rho ] 
\psi_\sigma 
+ {1 \over 4} \epsilon^{\mu\nu\rho\sigma} 
\bar\psi_\mu \gamma_\nu \gamma_5 [ \hat D_\rho, \hat D_\sigma ] 
\epsilon \nonu
\A = \A {1 \over 2} i e \bar\epsilon \gamma^\mu \psi_a 
\left( e_b{}^\nu \hat R_{\mu\nu}{}^{ab} 
- {1 \over 2} e_\mu{}^a \hat R \right), 
\label{rsvar3}
\ea
up to total derivative terms. In the last equality we have used 
\be
[ \hat D_\mu, \hat D_\nu ] \epsilon = 
{1 \over 4} \hat R_{\mu\nu}{}^{ab} \gamma_{ab} \epsilon 
\ee
and the properties (\ref{bilinearid}). Thus, the variations 
of ${\cal L}_{\rm E}$ and ${\cal L}_{\rm RS}$ 
cancel each other and the total Lagrangian (\ref{4daction}) is 
invariant under the local supertransformation (\ref{localsusy}) 
up to total derivative terms. 
\par
Next let us show the commutator algebra of two local 
supertransformations in eq.\ (\ref{commutatoralg}). 
We shall first consider the commutator applied on the vierbein 
\ba
[ \delta_Q(\epsilon_1), \delta_Q(\epsilon_2)] e_\mu{}^a 
\A = \A \delta_Q(\epsilon_1) 
\left( -i \bar\epsilon_2 \gamma^a \psi_\mu \right) 
- \left( 1 \leftrightarrow 2 \right) \nonu
\A = \A - i \bar\epsilon_2 \gamma^a \hat D_\mu \epsilon_1 
+ i \bar\epsilon_1 \gamma^a \hat D_\mu \epsilon_2 \nonu
\A = \A - i \hat D_\mu \left( \bar\epsilon_2 \gamma^a 
\epsilon_1 \right), 
\ea
where we have used the second equation in eq.\ (\ref{bilinearid}). 
Defining $\xi^\nu = i \bar\epsilon_2 \gamma^\nu \epsilon_1$ 
we obtain 
\ba
[ \delta_Q(\epsilon_1), \delta_Q(\epsilon_2)] e_\mu{}^a 
\A = \A - \hat D_\mu \left( \xi^\nu e_\nu{}^a \right) \nonu
\A = \A - \partial_\mu \xi^\nu e_\nu{}^a 
- \xi^\nu \hat D_\nu e_\mu 
- \xi^\nu \left( \hat D_\mu e_\nu{}^a - \hat D_\nu e_\mu{}^a 
\right) \nonu
\A = \A - \partial_\mu \xi^\nu e_\nu{}^a 
- \xi^\nu \partial_\nu e_\mu 
- \xi^\nu \hat\omega_\nu{}^a{}_b e_\mu{}^b 
- i \xi^\nu \bar\psi_\nu \gamma^a \psi_\mu, 
\ea
where we have used eq.\ (\ref{torsion}). 
This shows the last commutation relation in 
eq.\ (\ref{commutatoralg}) for $e_\mu{}^a$. 
Similarly, we can compute the commutation relation on the 
Rarita-Schwinger field. We need the supertransformation of 
the spin connection $\hat\omega_\mu{}^a{}_b$, which can be 
obtained by applying $\delta_Q$ on both sides of 
eq.\ (\ref{torsion}) as 
\be
\delta_Q(\epsilon) \hat\omega_{\mu ab} 
= {1 \over 2} \, i \left( \bar\epsilon \gamma_\mu \psi_{ab} 
- \bar\epsilon \gamma_a \psi_{b \mu} 
+ \bar\epsilon \gamma_b \psi_{a \mu} \right), 
\label{b12}
\ee
where $\psi_{\mu\nu} = \hat D_\mu \psi_\nu - \hat D_\nu \psi_\mu$. 
By using eqs.\ (\ref{b12}), (\ref{fierzid}) and (\ref{bilinearid}) 
we obtain 
\ba
[ \delta_Q(\epsilon_1), \delta_Q(\epsilon_2)] \psi_\mu 
\A = \A \left[ 
\delta_G(\xi) + \delta_L(\xi \cdot \hat\omega) 
+ \delta_Q(\xi \cdot \psi) \right] \psi_\mu \nonu
\A \A + {1 \over 16} \, \xi^\nu \left( 
\gamma_\nu {\cal R}_\mu + 2 \gamma_{\mu\nu\lambda} {\cal R}^\lambda 
\right) \nonu
\A \A + {1 \over 32} \, i \, \bar\epsilon_2 \gamma^{ab} \epsilon_1 
\left( 2 \gamma_{ab\mu\nu} {\cal R}^\nu 
- \gamma_{ab} {\cal R}_\mu -2 e_{\mu a} {\cal R}_b \right), 
\ea
where ${\cal R}^\nu = \gamma^{\nu\rho\sigma} \psi_{\rho\sigma}$. 
The field equation of the Rarita-Schwinger field is 
${\cal R}^\nu = 0$. 
Therefore, the commutator algebra closes only on-shell. 
\vspace{5mm}
%
%
\newcommand{\NP}[1]{{\it Nucl.\ Phys.\ }{\bf #1}}
\newcommand{\PL}[1]{{\it Phys.\ Lett.\ }{\bf #1}}
\newcommand{\CMP}[1]{{\it Commun.\ Math.\ Phys.\ }{\bf #1}}
\newcommand{\MPL}[1]{{\it Mod.\ Phys.\ Lett.\ }{\bf #1}}
\newcommand{\IJMP}[1]{{\it Int.\ J. Mod.\ Phys.\ }{\bf #1}}
\newcommand{\PR}[1]{{\it Phys.\ Rev.\ }{\bf #1}}
\newcommand{\PRL}[1]{{\it Phys.\ Rev.\ Lett.\ }{\bf #1}}
\newcommand{\PTP}[1]{{\it Prog.\ Theor.\ Phys.\ }{\bf #1}}
\newcommand{\PTPS}[1]{{\it Prog.\ Theor.\ Phys.\ Suppl.\ }{\bf #1}}
\newcommand{\AP}[1]{{\it Ann.\ Phys.\ }{\bf #1}}
\end{document}